\documentclass{aa}  
\usepackage{txfonts}
\usepackage{changes}
\usepackage{natbib}
\usepackage{pdflscape}
\usepackage{xcolor}	
\usepackage{longtable}
\usepackage{subcaption}
\usepackage{tabularx}

\usepackage{hyperref}
\hypersetup{colorlinks=true,linkcolor=blue,citecolor=blue,filecolor=blue,urlcolor=blue}

\hypersetup{colorlinks=true,linkcolor=blue,citecolor=blue,filecolor=blue,urlcolor=blue}

\newcommand{\Msun}{\ensuremath{\, {M}_\odot}}
\newcommand{\ocen}{$\omega$\,Cen}

\newcommand{\Thbb}{T$_{\rm HBB}$}

\begin{document} 

   \title{Noble gases Neon and argon: a role for the chemical patterns of multiple populations in globular clusters?}
   \authorrunning{P. Ventura et al.}
   \titlerunning{Updating the AGB model for the formation of second generation}

   \author{ P. Ventura\inst{1}, F. D'Antona\inst{1}, M. Tailo\inst{2}, 
    A. P. Milone\inst{3,4}, F. Dell'Agli\inst{1}, A. F. Marino \inst{3}
          }

   \institute{
              Istituto Nazionale di Astrofisica -- Osservatorio Astronomico di Roma, Via Frascati 33, Monte Porzio Catone, I00078
              \and 
              Dipartimento di Fisica e Astronomia Augusto Righi, Università degli Studi di Bologna, Via Gobetti 93/2, 40129 Bologna, Italy
              \and Istituto Nazionale di Astrofisica -- Osservatorio Astronomico di Padova,
              Vicolo dell'Osservatorio 5, Padova, I-35122
              \and Dipartimento di Fisica e Astronomia ``Galileo Galilei'', Univ. di Padova, Vicolo dell'Osservatorio 3, Padova, I-35122
             }


 
  \abstract
  {The role of the initial elemental abundances in the nuclosynthesis of massive asymptotic giant branch (AGB) stars  and second generation abundances in Globular Clusters.} 
  {We look at the difficulties of the AGB models into fully reproducing the abundance patterns in the second generation stars of Globular Clusters, and focus on the sodium destruction in models reaching the high hot bottom burning temperatures needed to efficiently cycle oxygen to nitrogen.}
 {We build AGB models at the nominal [Fe/H] of the cluster NGC\,2808, increasing the initial neon abundance by a factor 2--4 with respect to the `standard' abundances obtained by scaling the solar values down to the metallicity of this cluster, and explore the average abundances in the ejecta obtained by adopting smaller mass-loss rates. }
 {Higher neon produces higher sodium in the AGB envelope. Lowering  the mass-loss rate allows both to keep reasonably large sodium abundances and to increase the depletion of oxygen and magnesium. A balance between the lower mass-loss rates and the necessity of not increasing too much the episodes of third dredge up gives a neon abundance larger by a factor two and a mass-loss rate smaller by a factor four as best compromise. The comparison with the abundances in NGC\,2808 shows a better agreement than the standard models for all the patterns of abundances, but the extreme stars (group E) can not be explained in this way. 
 Models of initial mass $\geqslant 6.5$\Msun for the same chemistry (super--AGBs) are closer to the E data, but the oxygen depletion remains too small. We also compute models slightly less rich in iron, and show that the super--AGB ejecta composition becomes more compatible with the data.  
Thus, we propose that the extreme population in NGC\,2808 is composed of stars having a slightly smaller metallicity, and sketch a possible  scenario for its formation, in the framework of the hierarchical clusters assembly scenario. Finally, abundances of potassium are larger by $\sim$0.2\,dex in the E group. The explanation in terms of burning of the initial argon requires a drastic increase of the relevant cross section, similar to the adjustment required in the exploratory models for the cluster NGC\,2419, but it becomes at least easier when we assume a larger initial argon abundance.}
 {The abundances of the noble gases neon and argon at low metallicities may be an important tool to better reproduce the abundances of light elements in the framework of the AGB model for globular clusters. The larger Ne and Ar abundances we need may either be a plain consequence of the uncertainties in the solar abundances of these elements or in part arise from an incomplete summing up of the sources for their nucleosynthesis at low metallicity. Thus we launch a plea for a better understanding of their galactic chemical evolution of noble
 gases.}


   \maketitle
%

\section{Introduction}
\label{intro}
Our knowledge about the evolution of the abundances of elements in the Universe comes first of all from studying the chemical composition in the atmospheres of stars of different metallicities, relating it to the chemistry of the forming gas. Observing stars of different ages and chemistry, we can build a general framework to understand the chemical evolution of elements throughout the lifetime of the Universe, taking into account the nucleosynthesis in supernovae and other elements factories such as the s--process sites.  The main standard reference along this road is the Sun, whose abundances come in part from the study of the solar photospheric spectrum\footnote{Abundances in the spectrum are not the same as the abundances at  formation in the solar nebula, as models must also take into account early nuclear processing during the phases in which the Sun was fully convective, and the action of gravitational settling at the bases of the main sequence convection zone \citep[see, e.g.][]{lodders2003}.} and in part from the analysis of meteorites. Atmospheric solar abundances agree within 10\% with those in the CI-type Carbonaceous meteorites, when normalized to the same scale \citep{lodders2003}.  Unfortunately, meteorites are depleted in the elements which readily form gaseous compounds, like H, C, N, O and the noble gases.
For these elements we have to rely only upon the atmospheric abundances. Concerning neon, in the solar corona we measure its abundance ratio to oxygen or magnesium, while the abundance of these latter elements are determined  from the atmospheric lines. This introduces an uncertainty, a problem fully appreciated when the revision of the solar abundances \citep{asplund2006} provoked a disagreement of the helioseismological determinations of the solar properties, mainly the depth of the convective solar envelope and the inferred surface abundance of helium, \citep{bahcall2005a}, so far very well matched by the previously adopted abundances by \cite{grevessesauval1998}. Among many models explored to get back a good match of the helioseismic solar parameters,  \cite{bahcall2005b} proposed to rise the neon abundance by 0.4-0.6\,dex. Along this same direction,  \cite{draketesta2005} derived a number ratio Ne/O=0.41 directly from the fluxes of O and Ne in the X--ray spectra of the coronae of cool stars within 100 pc, and this value is a factor 2.5 larger than the number ratio Ne/O=0.15 of the solar abundances.
In the more recent research on the solar abundances, oxygen values are not as low as in the \cite{asplund2006} determination  \citep[see, e.g.][]{grevesse2011}, but helioseismology still requires adjustments to correctly reproduce the Sun. The approach is now to explore several different uncertain parameters  \citep[e.g.][]{villante2020}, and a factor $\sim$2 increase in the {\bf neon} abundance is still required for a good fit. Another constraints from the seismic inversion comes from the synthesis of the adiabatic exponent $\Gamma_1$\ profile, quite sensitive to the heavy elements ionization. \cite{baturin2024} derived in this way for the solar convection zone an oxygen and carbon abundances consistent with the values by \cite{asplund2021}, and a neon abundance  $\sim$20\% larger.
Other abundance adjustments came out in recent years: the ratio Ne/O in the transition region of the quiet Sun was increased by \cite{young2018} from the previous ratio, implying an increase by 40\% \citep[with respect to ][]{asplund2009} in the photospheric abundance of neon. An updated discussion on the uncertainties is given in \cite{asplund2021} and \cite{lodders25}.

We shift our attention to the abundances of the stars in globular clusters (GCs). Two main parameters characterize their chemical composition:  the `metallicity' (and mainly the iron content, which testifies the chemical enrichment by supernovae) and  the $\alpha$-elements abundances relative to iron \citep[larger than solar in the low metallicity environment and in GCs, as expected based on the prevalence of core collapse supernovae nucleosynthesis during the first phases, see e.g.][]{matteucci1986}. \\
Historically described as `single stellar populations', made up of stars sharing a single age and the same chemical composition, GCs hide multiple populations:
in all clusters (exceptions are very rare until today), a majority of stars shows evidence for matter processed at high temperature by p--capture reactions, the rest is similar to field stars of the same metallicity. This apparently simple subdivision hides the complexity of the problem.  Reasonably assuming that the p-capture reactions have taken place in a previous stellar environment, named `first generation' (1G), it results that 60-80\% of the stars are `second generation' (2G). Thus the 1G is deceiving us, as we do no longer see it in the same place, either because it belonged to a much wider system than the cluster (e.g. for a cluster forming in a dwarf galaxy now dispersed) or because the majority of the 1G stars escaped the cluster. In addition, there is a great variety in the number of populations and in the complexity of chemical patterns (some clusters also display multiplicity in iron abundance) and any model devised should allow for this complexity in a natural way. As a general reference to the framework of multiple populations see, e.g. the recent review papers by \citet{gratton2019, milonemarino2022} and the vast literature quoted therein. In conclusion, in GCs we are facing a very rapid chemical evolution which took place in the very first phases of life of globular clusters, and this requires pollution modalities very different from the chemical evolution models, which account for the secular evolution of the elements along the lifetime of the Galaxy, e.g. the increase in iron and other elements from the population II to the solar abundances (and beyond). The latter models calibrate parameters and  sometimes adopt educated guesses on the influence of different polluters, to reproduce the galactic abundance trends.
\\ 
No model is able to fully account for the chemistry of 2G stars. In particular, the AGB model \citep[originally presented in its modern version in][]{ventura2001, dercole2008} has been explored in much more detail than the others and is obviously the model most subject to criticism, allowed by its more detailed predictions. 
In spite of its fundamental advantages, not shared by any of the other models proposed so far, chemistry is the major problem of the AGB model, because it is unable to provide a full explanation of the O--Na anticorrelation. \cite{renzini2015} examine in detail how the relative roles of the cross sections of  p-captures on oxygen (fluorine) and on sodium (to neon) are difficult to reconcile with the need for strong oxygen depletion and sodium survival. \\
The most extreme stars in clusters like the prototype NGC\,2808 exhibit a depletion of the
surface oxygen by more than a decade with respect to first generation stars, so both burning temperatures and timescales must be large/long enough to convert a great fraction of oxygen into fluorine, but  p-captures deplete the sodium too, while observations show that these stars preserve a sodium overabundance. \\
In this paper we explore the possible role of the initial neon abundance on the chemical composition of the ejecta of massive AGB and super--AGB stars to understand whether this input can improve the comparison with the composition found in 2G stars. 
Our starting point follows the idea that the solar {\bf neon} abundance is underestimated by a factor from 2 to 4, so that the neon abundance at the smaller metallicities of GCs is also underestimated by the same factor. The neon increase we propose is additional to the $\alpha$-enhancement factor already taken into account when scaling the neon abundances. In the end, we will consider a factor two increase as the best compromise.
A different interpretation of this approach is to consider the solar neon abundance correct, and anyway assume that neon,  at lower metallicities, was indeed more abundant than suggested by scaling. Unfortunately, the scarce present day neon abundance determination in planetary nebulae (see Sect. \ref{PNNeon} and Fig.\,\ref{figne}) do not help to constrain the more plausible choice. 
\\
In Sect.\ref{agb} we shortly outline the main accomplishments, and and in Sect.\ref{whyneon} we remind the main difficulty of the AGB model to account for the chemistry of the 2G stars; 
in Sect.\,\ref{neonmassloss} we present models with reduced mass-loss rate and enhanced neon aimed at explaining the different populations found in NGC\,2808, one of the GCs showing the most extreme light elements anticorrelations, well studied for many different elements; we also explore models with slightly smaller iron content, and show that their ejecta composition match better the abundances in the most extreme stars of the cluster;
in Sect.\,\ref{clumps} we propose a formation model for NGC\,2808 which may account for the extension of all the anticorrelations.
In Sect.\,\ref{pot} we discuss preliminary models to synthetize potassium, and in  Sect.\,\ref{PNNeon} we shortly discuss the neon abundances in planetary nebulae.
Although the results presented in this work go along the right direction, solving of the several open problems requires careful adjustments of the inputs.

\begin{figure}
\vskip -40pt  
\begin{minipage}{0.48\textwidth}
\resizebox{1.\hsize}{!}{\includegraphics{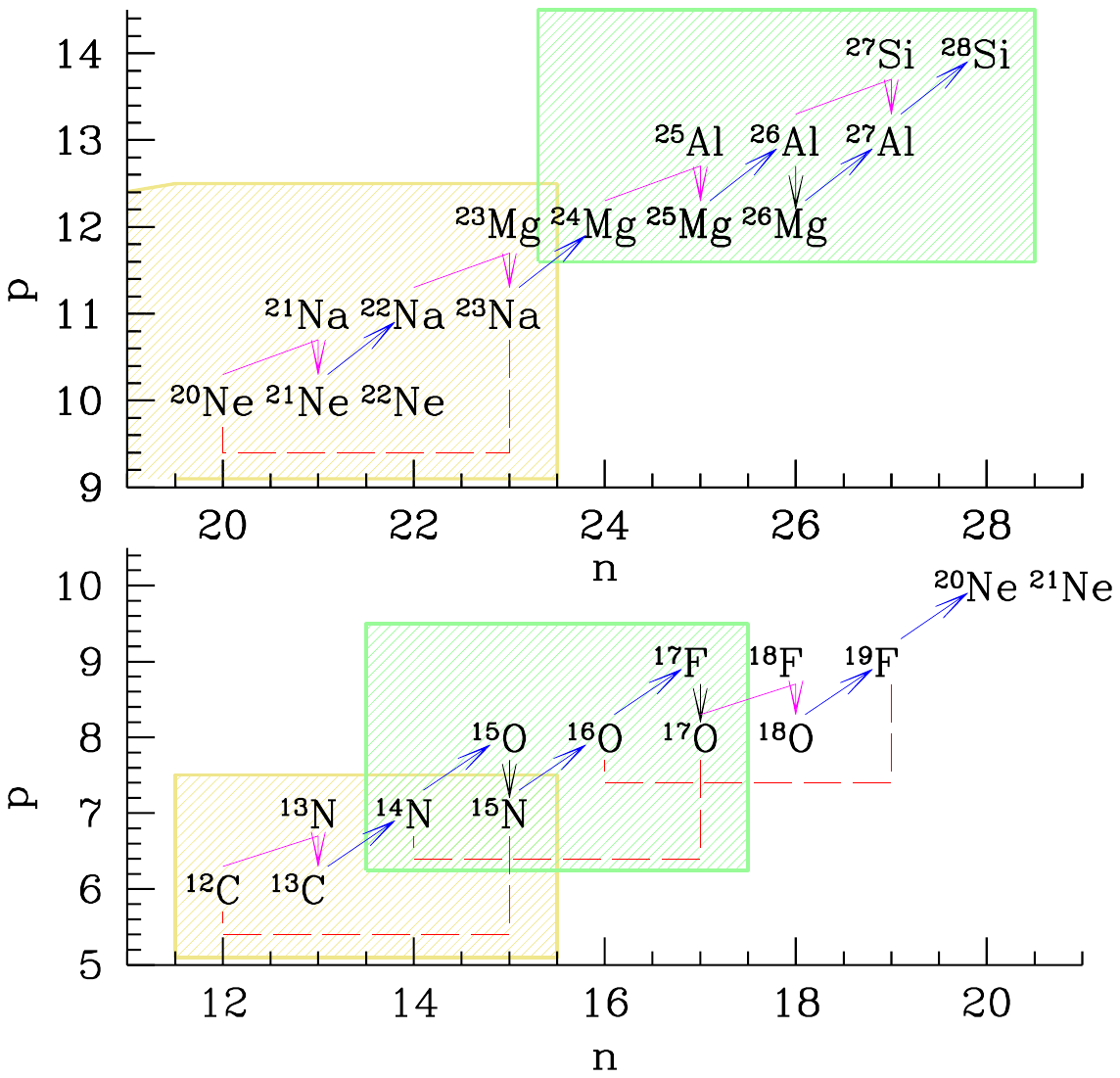} }	 
\end{minipage}
\vskip -70pt  
\caption{Schematic p--captures for the CN (yellow box) and NO (green box) cycles (bottom panel) and for the Ne--Na (yellow) and Mg--Al and Al--Si (green) cycles (top panel). Blue arrows indicate $\rm (p,\gamma)$
reactions, pink trajectories indicate p-captures followed by $\beta$ decays,  dashed red lines show (p, $\alpha$) reactions. The processing occurs for increasing hot bottom burning temperatures.
}
\label{fig:1}       
\end{figure}

\section{The AGB model}
\label{agb}
Although many factors related to a cluster internal dynamics and its complex interplay with the external galactic environment (particularly in the early stages of galaxy formation) have certainly played a key role in determining the current properties of multiple-population clusters, any effort aimed at identifying the fundamental ingredients in the formation of multiple populations should be driven by the strong observational constraints coming from the chemical abundance properties and patterns revealed by spectroscopic and photometric observations. We need to identify which are the fundamental clues on the possible sources of gas out of which 2G stars formed. In very short and minimal fashion, the strongest chemistry constraints we must deal with are: 
\begin{enumerate}
\item The inner content of helium of 2G stars is larger ---generally moderately larger--- than the helium content of the 1G, but it appears to be limited to a maximum mass fraction Y$\simeq$0.35--0.38, with such large abundances found only in a few clusters;
\item The typical anticorrelations found in the abundances of p--capture elements (C-N, N-O, Na-O and Na-Al)  are common in all clusters; some clusters also show Mg-Al and Mg-Si anticorrelations;
\item Very few clusters shows a potassium vs. magnesium anticorrelation, and possibly a mild correlation K-Ca. 
\end{enumerate}
The increase in helium content is to be expected, as we are dealing with hydrogen burning, and the existence of an upper limit is an important clue. Higher and higher temperatures are required 
{\bf to account for} the p-captures which are at the basis of the different anti correlations. The main cycles of p--captures are shown in Fig.\,\ref{fig:1}, where the temperatures necessary for the activation of the cycles are $\sim$20\,MK for the CN cycle (yellow box in the bottom panel), $\sim$40\,MK for the NO cycle (green box). 
The $^{24}$Mg, that had been {\bf manufactured} by He-burning reactions in the {\bf massive star progenitors} of the cloud forming the 1G stars, can be subject to p-captures at the base of the convective envelope only for temperatures larger than 100\,MK, and in a few cases the p-capture chain may extend to Al, increasing Si (see Fig.~\ref{fig:1}). Temperatures as high as 100\,MK are not reached in the centre of regular massive stars\footnote{The temperatures required to process $^{24}$Mg by p-captures allow to dismiss the rotating massive stars and
massive binaries as pollutants for the formation of the 2G.}. This led to the selection of supermassive stars as a possible processing site \citep{denissenkov2014}. In spite of the high temperature nuclear processing in the gas now forming the 2G stars, the fragile element lithium is preserved, with abundances close to standard, or is only partially depleted in stars showing the most extreme anticorrelations.
\\
In the models most extensively studied in detail, p-captures do not take place in the stellar cores, but at the basis of the convective envelopes of massive Asymptotic Giant Branch (AGB) stars, where the temperature at the bottom (\Thbb) becomes so high that the elements are subject to `hot  bottom burning' \citep[HBB,][]{bs1993}. During the AGB evolution, the processed gas is transported by turbulence to the surface of the AGB, and is lost into the outer space by stellar winds and planetary nebula ejection. 
This phase is short-lived, and the temperatures of HBB depend on the model adopted to describe convection. However, the most massive and very low metallicity models that employ highly efficient convection may achieve extremely high temperatures (120-130 MK). In these conditions, p-captures on argon may occur, which may explain the potassium overabundances reported above \citep{ventura2012ngc2419}.

O--depletion in the massive AGB envelopes of low metallicity stars subject to HBB was first extensively found in the models built by \citet{ventura2001}. At the epoch, no other stellar modellers confirmed these results. Two main physical inputs were the reason for the big difference in the new results: 
\begin{itemize}
\item The convection model: all the other models employed the standard mixing--length theory \citep{bohmvitense1958}, and the efficiency parameter was generally left at the value obtained (at that time) to fit the Solar model. The new models by Ventura et al. (2001) employed the Full Spectrum of Turbulence (FST) model \citep{cm1991, cgm1996}. This choice resulted in a higher convection efficiency at the bottom of the convective envelope of AGB models, producing two main effects: 1) a shorter distance between the outer edge of the CNO-burning shell and the base of the convective envelope, which increased the efficiency of the HBB p–capture reactions. The products of these reactions were transported throughout the convective envelope and finally ejected to the interstellar medium; 2) an extra-luminosity, triggering a faster growth of the stellar luminosity and hence of mass loss rates \citep{ventura2005a}.
\item The mass-loss treatment: we adopted the \cite{blocker1995} formulation for the mass-loss rate
\begin{equation}
\dot{M}=4.83 \cdot 10^{-9} M^{-2.1} L^{2.7} \dot{M}_{R}
\label{eq1}
\end{equation}

where $M$\ and $L$\ are the stellar mass and luminosity, and $\dot{M}_{R}$ is Reimer’s mass loss rate, depending on the stellar 
radius $R$ (all given in solar units):

\begin{equation}
\dot{M}_R= 4\cdot 10^{-13} \eta { L R \over M}
\label{eq2}
\end{equation}
The parameter $\eta$\ in our models has been calibrated on the statistics of Lithium rich giants in the Magellanic Clouds \citep{ventura2000lmc} to $\eta=0.01 - 0.02$. 
Such a mass-loss rate, dependent on a high power of the luminosity, together with the effect of the efficient convection, reduce the overall duration of the AGB evolution by a factor of $\sim$3 with respect to solar calibrated MLT computations\footnote{\bf use of other frequently used prescriptions for mass-loss, such as those proposed by \citet{vw93} and
\citet{jacco}, would diminish this difference to a factor 
$\sim 2-2.5$}, thus enhancing the mass loss during the phases of maximum Na production, and reducing the mass lost when Na was depleted in the envelope \citep{ventura2005b}. 
In addition, they selected the most favourable range allowed for the cross sections of the NeNa-cycle reactions (considering the existing uncertainties), and maintained the Na abundances that conform to the observational data \citep{vd2006, ventura2009}.

\end{itemize} 

In summary, by choosing the most efficient convection model available (the FST), by tuning the mass loss rate to values compatible with the analysis of the lithium-rich giants in the Magellanic Clouds, and by tuning the cross sections of the NeNa-cycle reactions, the ejecta of massive AGB stars showed both oxygen depletion and preservation of sodium at levels compatible with the observed Na–O anticorrelation.

The AGB model is able to account for a good number of features of GCs. In particular, stellar models and dynamical models have been studied in detail and provide a reasonable match to the complexity and variety of chemical patterns in the multiple populations. The long evolutionary times of AGB masses give long timescales for the formation of the 2G and easily account for different chemical pollution patterns, following the different structures of AGB of different mass \citep{dantona2016}. \\

\section{Why the most successful model is unsuccessful?}
\label{whyneon}

\begin{figure}[t]
\vskip -40pt  
\begin{minipage}{0.98\textwidth}
\resizebox{.5\hsize}{!}{\includegraphics{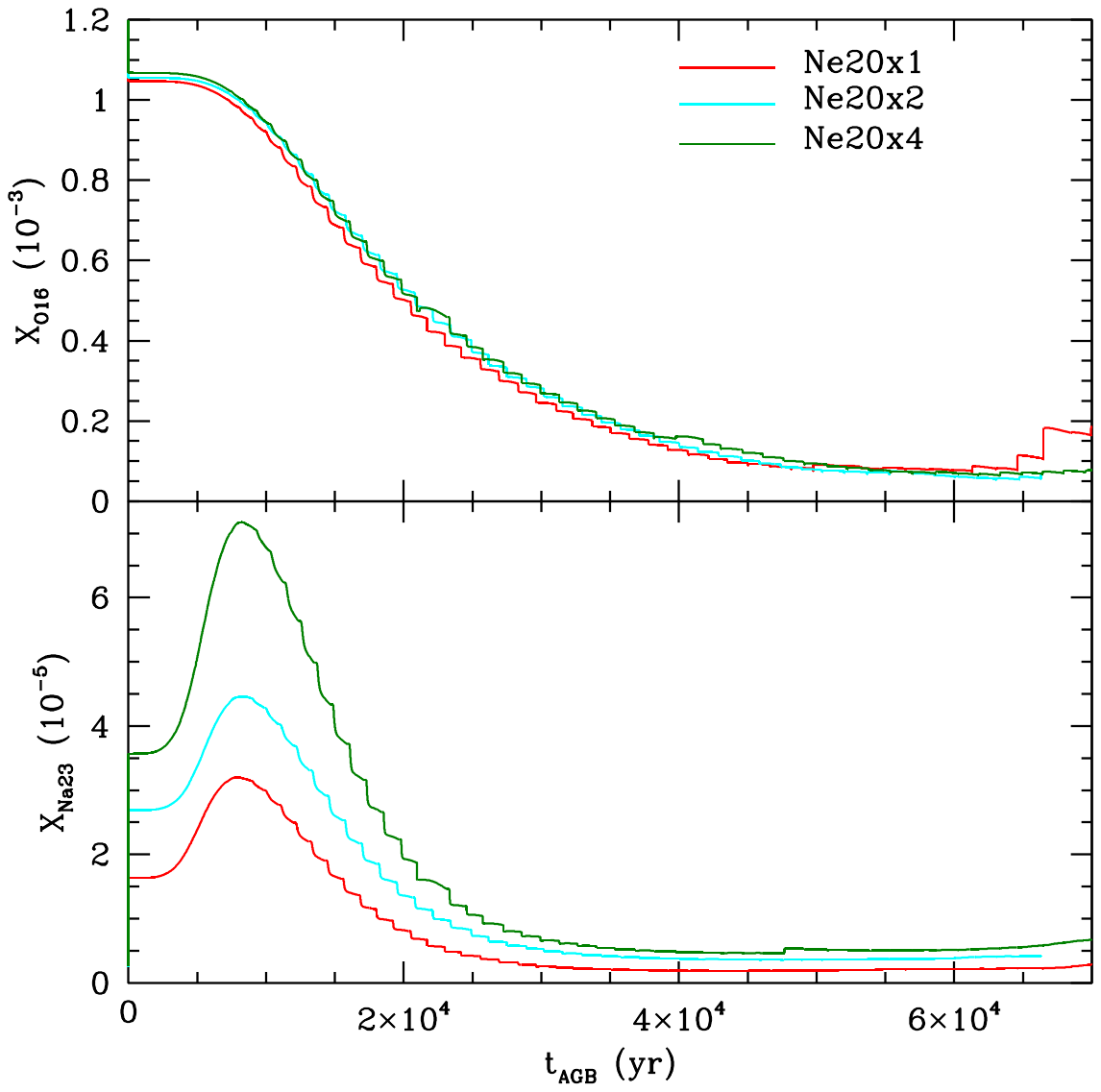} }	 
\end{minipage}
\vskip -70pt  
\caption{Changes of the sodium abundance versus time for the 6\Msun\  AGB evolution, when we vary the Ne20 initial abundance in the models. For simplicity, we show models with the same mass-loss rate calibration $\eta$=0.002, that is 1/10 of the standard value $\eta$=0.02.}
\label{fig:2nao}       
\end{figure}

The most critical aspect of the AGB model is illustrated in Fig.\,\ref{fig:2nao}, which shows the evolution of the surface mass fractions of sodium and oxygen with time, along the AGB phase of $\rm 6~M_{\odot}$ model stars, calculated with different initial neon abundances. The depletion of sodium necessarily accompanies the depletion of oxygen during the AGB evolution, because the cross section for p-captures on $^{23}$Na nuclei is larger than the cross section for p-captures on $^{16}$O nuclei \citep[see, e.g.][for a discussion]{renzini2015}. Thus, when oxygen burns, sodium also burns. Sodium may preserve an average abundance higher than in the first generation only because the second dredge-up has replenished the envelope with sodium and neon from the helium intershell, and neon has been rapidly converted to sodium thanks to the high \Thbb.
Even if sodium and oxygen are both burned in the following AGB evolution, the average abundance of sodium in the ejecta may remain large. 

\subsection{The comparison cluster: NGC\,2808}
All the comparisons are made with the abundance patterns in NGC\,2808. The choice is well motivated but also needs justification. It is a standard result of AGB evolution that the temperature (\Thbb) at the base of the convective envelope for a given mass increases for lower metallicities, corresponding to lower opacities of the envelope matter. Stars in NGC\,2808 have a relatively high metallicity \citep[{[Fe/H]}=--1.1,][]{carretta2015}, but, in spite of this, they display very extended abundance patterns, so NGC\,2808 is an important benchmark for any new modelling. In addition, the stars also show a spread by a factor $\sim$2 in the abundances of potassium \citep{mucciarelli2015}, although this element is seen to vary mainly (by an entire decade) only in the very low metallicity cluster NGC\,2419 \citep{cohenkirby2012}\footnote{Note that a potassium -- magnesium anticorrelation was found  only in NGC 2419 and $\Omega$ Centauri, so the extreme stars in NGC\,2808 represent indeed a very peculiar sample.}. Although it was not revealed in the spectroscopic analysis, it is possible that the 1G stars of NGC\,2808 display a moderate metal inhomogeneity, $\delta$[Fe/H]=0.25$\pm$0.06\,dex \citep{lardo2022}.  \\
The accurate catalogue of abundances in \cite{carretta2015}, subdivided into five subgroups of increasing chemical ``anomalies" constitutes the most complete  spectroscopic benchmark we must comply with, and are
 shown in Figs. \ref{fig:3} and  \ref{fig:4}. The stars were assigned to the  P1, P2, I1, I2 and E classes based on their location in the Mg--Na plot, where the groups are well distinguished. The discreteness of the groups is a significant hint that the formation of the multiple populations occurred in bursts, and probably from matter contaminated by polluters in which the gas was processed by p--captures at different temperatures. In the AGB model, this is a very probable situation \citep{dantona2016}, because the second  generation formation occurs on a long timescale of several tens of million years, as modelled by \citep{dercole2008, dercole2016, calura2019}. We can take this into account when we compare the data with the new exploratory models.

\subsection{Abundances in the standard models}
In our first comparisons with the abundance patterns of NGC\,2808 we used models with a metals mass fraction Z=0.001, where the elemental abundance ratios followed \cite{grevessesauval1998}
abundance distribution. In the following, the solar abundances were revised, and it became clear that the HBB depends on the opacity of the envelope, dictated by the iron abundance more than by the total metallicity. It is necessary then to update our comparisons, taking into account more strictly the specific abundances and abundance ratios. In the present models, our standard mixture follows \cite{asplund2009}, where the percentage of neon is 0.094 and Z$_\sun$=0.014. The solar mass fraction of neon is then Z$_{\rm Ne}=1.31\times 10^{-3}$. 
For NGC\,2808 we adopt  [$\alpha$/Fe]=0.4, so that the standard neon fraction becomes 0.1208 of the total metallicity, fixed at Z=2.3$\times 10^{-3}$. Thus the neon mass fraction in the standard models is 2.78$\times 10^{-4}$. When we define models with a factor x2 or x4 neon, we mean this mass fraction is multiplied by 2 or 4, while the other abundances remain the same (in particular [Fe/H]). With this choice, the total metallicity  Z increases, and obviously also the fractional values of the other abundances readjust.

\subsection{Standard models do not deplete enough oxygen}
\label{oxygen}
\begin{figure*}
\vskip -30pt  
\begin{minipage}{0.95\textwidth}
\resizebox{1.\hsize}{!}{\includegraphics{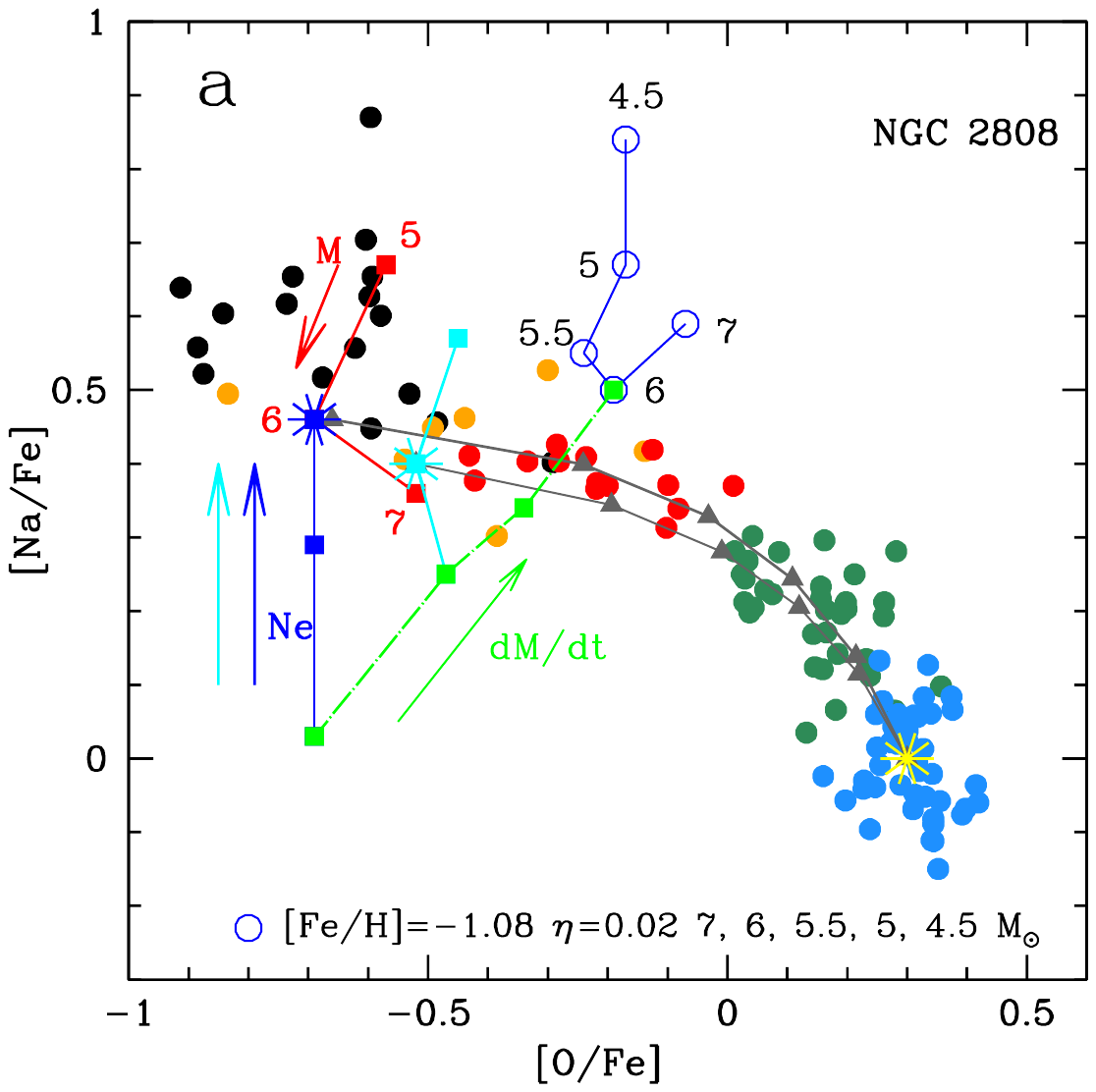}\includegraphics{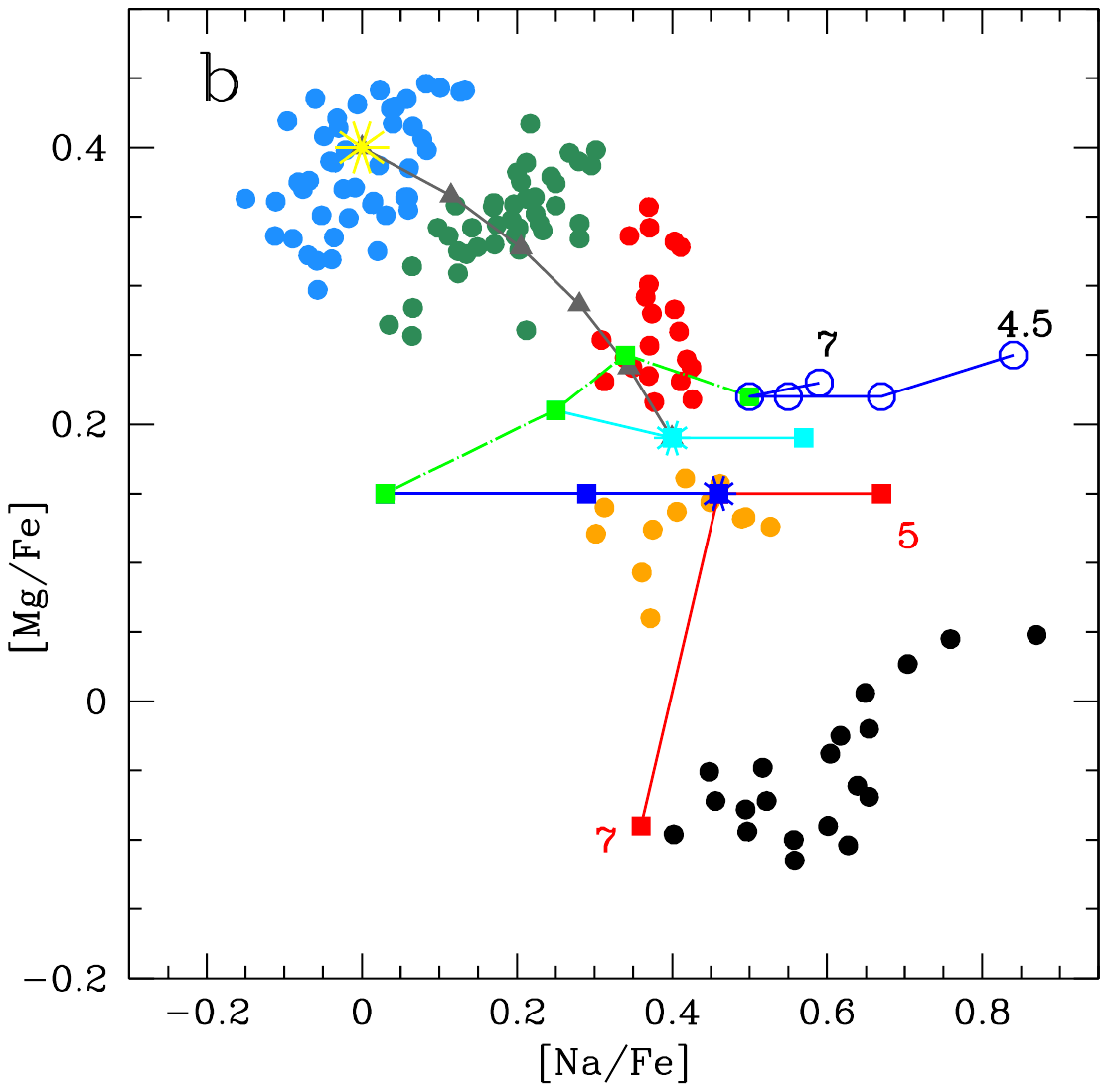}\includegraphics{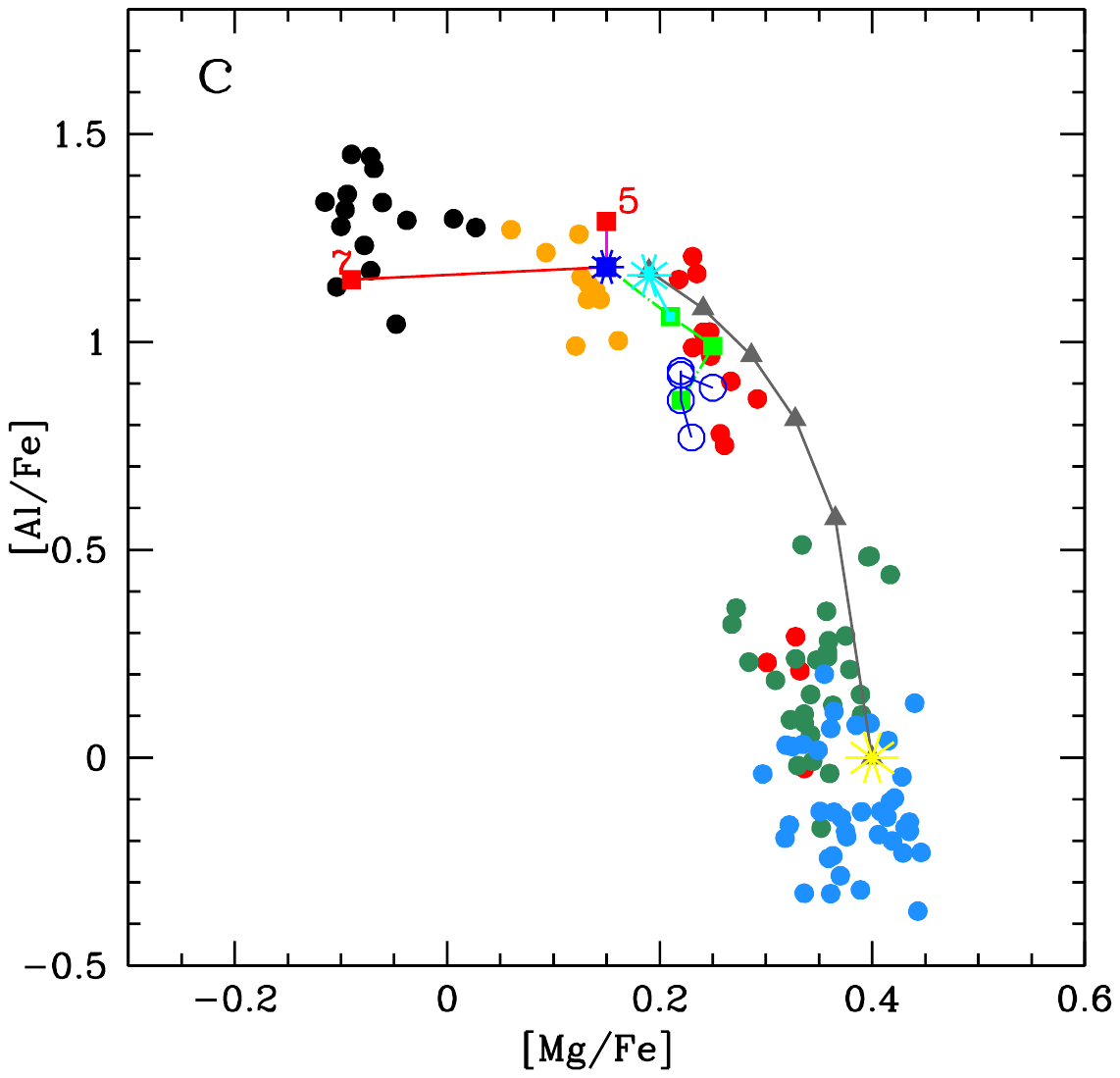}}	
\vskip -70pt
\resizebox{1.\hsize}{!}{\includegraphics{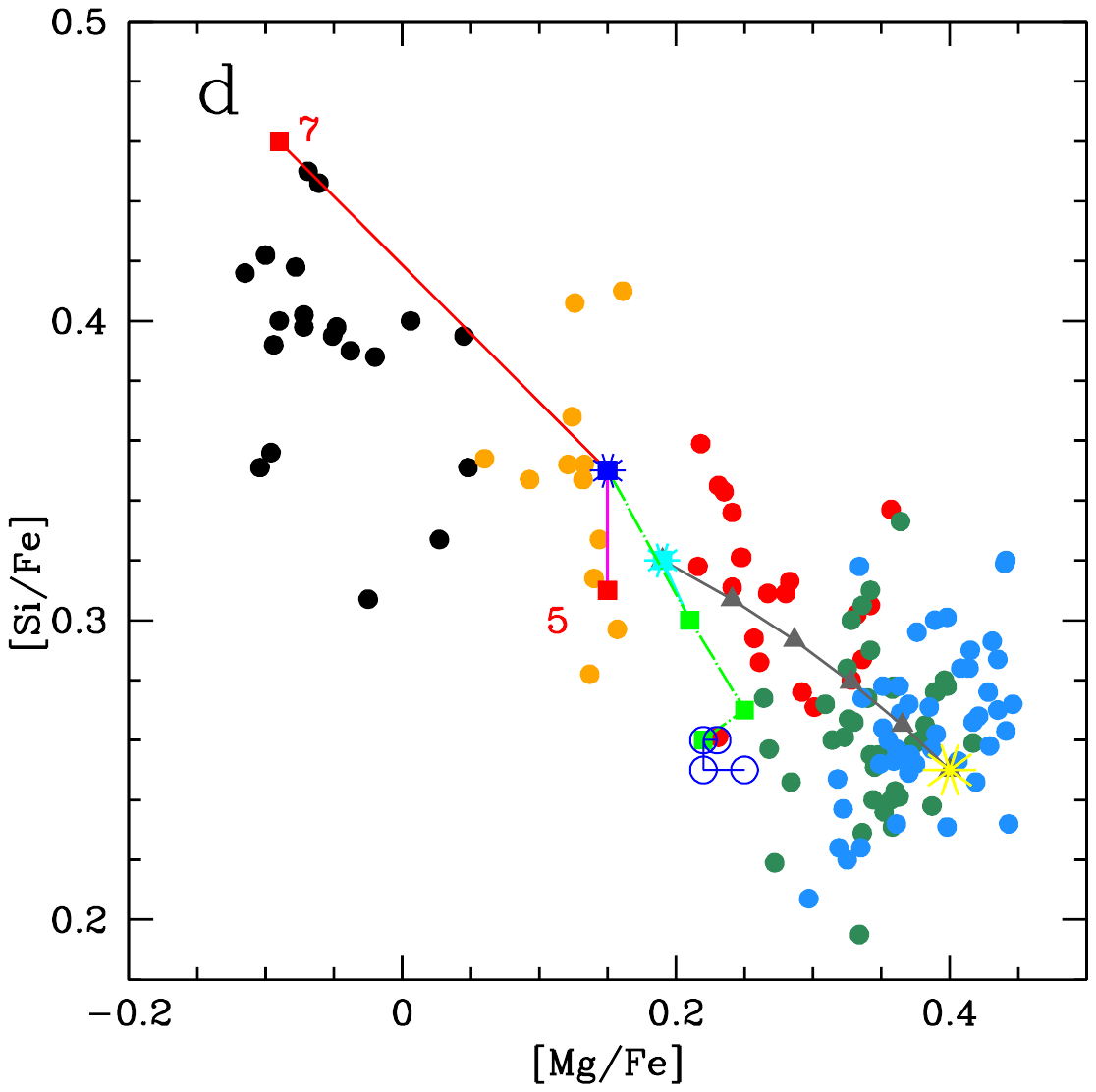}\includegraphics{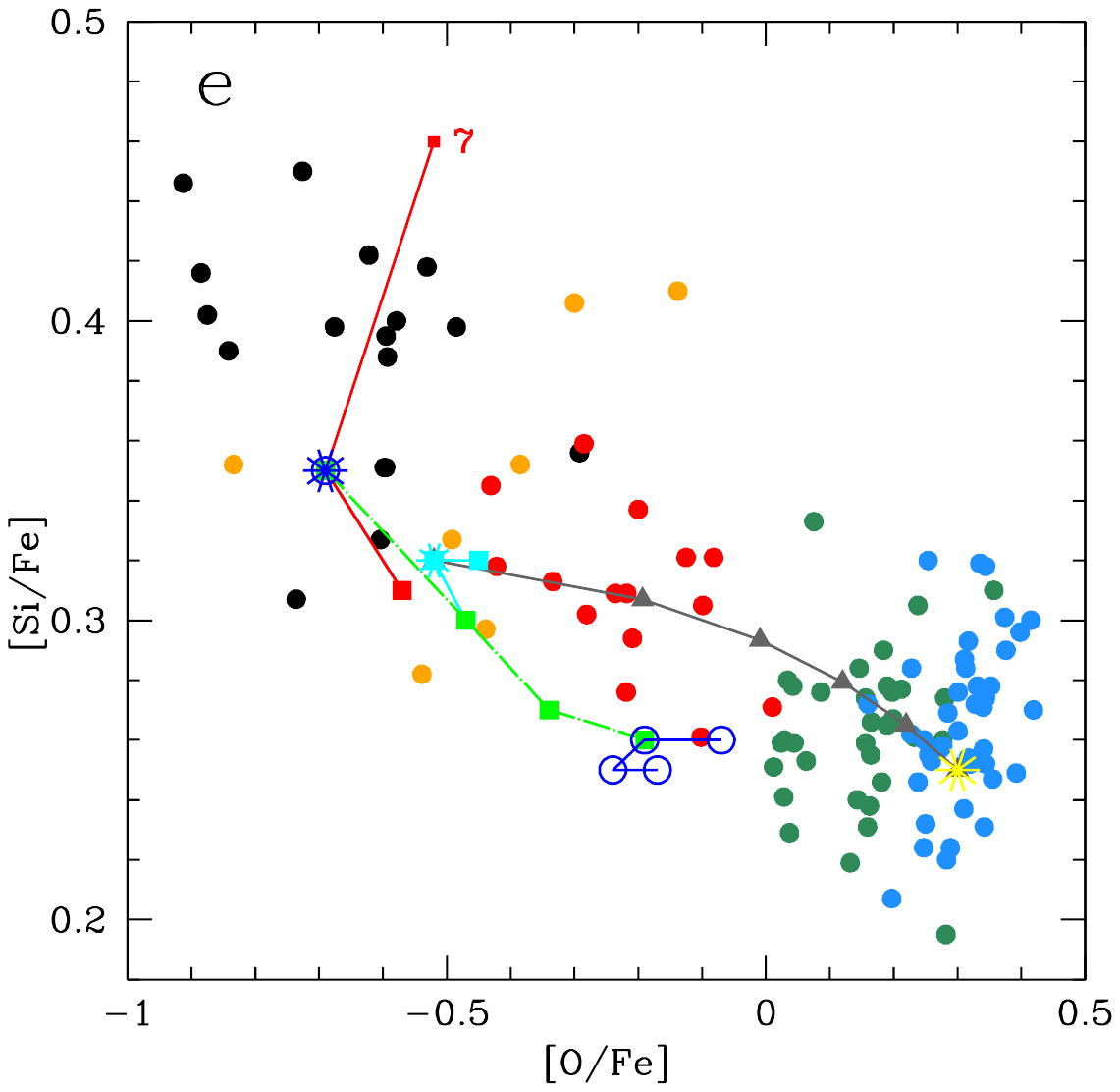}\includegraphics{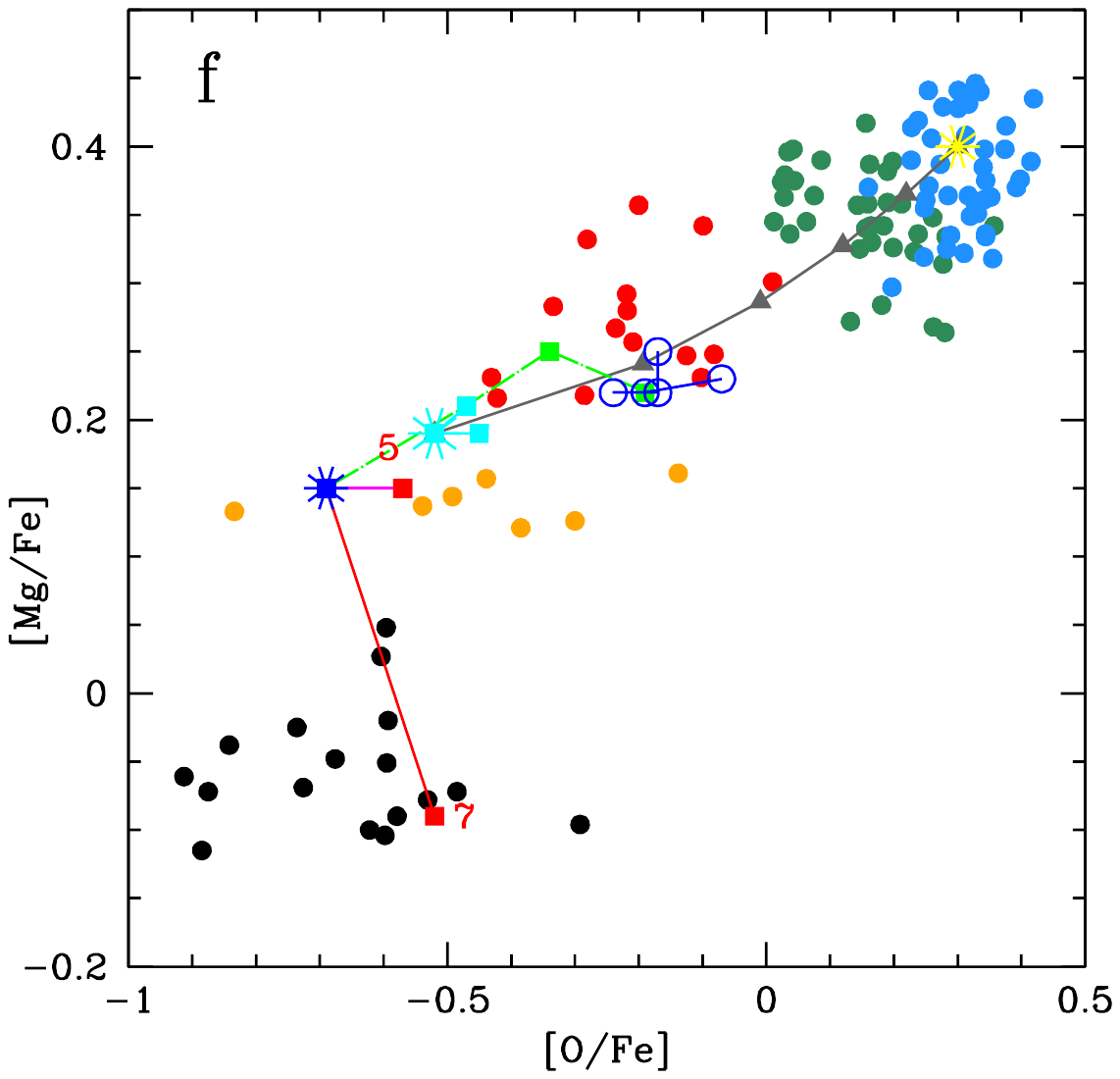}}	
\end{minipage}
\vskip-50pt
\caption{ Anticorrelations of p--capture elements in the  GC NGC\,2808 \citep{carretta2015}. The diagram of sodium--magnesium data (panel b) was the basis for the subdivision into different sub-classes: P1 (blue), P2 (green), I1 (red), I2 (orange) and E (black). All panels show: 1) standard average abundances in the AGB ejecta by for a metallicity Z=2.3$\times10^{-3}$, [$\alpha$/Fe]=0.4, e.g. [Fe/H]=--1.08 (open blue circles); 2) exploratory models (colored squares and connecting lines) with the same initial composition and identified by: GREEN: ejecta of the 6\Msun, with standard neon and decreasing mass-loss rates (models $\eta x$Ne1). Starting from the standard abundances for $\eta$=0.02, we plot  $\eta$=0.01 (1/2 standard), 0.005 (1/4 standard) and 0.002 (1/10 standard); BLUE: 6\Msun for $\eta$=1/10 standard, with increasing Neon abundance from standard (solar scaled, for $\alpha/Fe$=0.4) to 2 and 4 times standard (models $\eta$\,0.1Ne4); RED: abundances for the 5 and 7\Msun\ for $\eta$\,0.1Ne4; CYAN: starting from mass loss 1/4 standard, and  increasing Neon at 2 and 4 times the standard abundance  (models $\eta$\,0.25Ne2 and $\eta$\,0.25Ne4). The blue and cyan big stars define the location of the models having sodium consistent with the O--Na patterns and discussed in the text: $\eta$\,0.1Ne4 and $\eta$\,0.25Ne2, which we will consider also in the other panels.
The dark grey lines with triangles define dilution of the 6\Msun location $\eta$\,0.25Ne4  with increasing quantities (20, 40, 60 and 80\%) of gas with standard composition, identified by the yellow star. 
}
\label{fig:3}       
\end{figure*}

The Na and O average abundances in the ejecta for different AGB masses are seen as blue open circles in panel a of Fig.\,\ref{fig:3}.  The results refer to models of [Fe/H]=--1.08 [$\alpha$/Fe]=0.4 for a total metallicity Z=2.3$\times 10^{-3}$.
The details of the input physics of the models are given in \cite{dellagli2018}, we point out that the mass-loss rate follows \cite{blocker1995} formulation (see eq.\,\ref{eq1} and \ref{eq2}) with $\eta$=0.02. 
 The iron content of these models is very close to the value determined for NGC\,2808, whose abundance patterns are displayed in the panels of the figure.
We see that oxygen in the models does not match the I2 nor the E group, as its maximum depletion, starting from an initial abundance [O/Fe]=0.4,  is $\sim$0.55\,dex, for the 5.5\Msun. 
Previous comparisons with NGC\,2808 \citep[e.g.][]{dantona2016} were based on models by \cite{ventura2013} computed for a lower metallicity (Z=$10^{-3}$), for which  p-processing is stronger and therefore shows a greater oxygen depletion ($\sim$0.8\,dex for the 6\Msun models), although not sufficient to match the  E subclass. \\
It is important to spend a few words on the typical "hook" shape of the O--Na ejecta abundances as a function of the mass. The average abundances in the ejecta depend on the interplay between two main parameters of the models: \Thbb and the mass-loss rate. Oxygen burning will be more effective the higher is \Thbb\ and the lower is the mass-loss rate. Indeed both these quantities increase with the evolving mass: higher mass models have larger luminosities, larger mass-loss rates and consequently shorter AGB evolution, so that oxygen has less time to be converted into nitrogen by HBB. So the oxygen depletion will first increase for decreasing mass, until the lower \Thbb\ prevails and oxygen increases again in the ejecta. There will be a `turning mass', typically in the $\rm 5-6~M_{\odot}$ range \citep{dellagli2018}, for which a minimum oxygen abundance is reached, and it will be higher for lower metallicity, having higher  \Thbb. Lower mass-loss rates for the same physical inputs produce a higher oxygen depletion. \\
The sodium abundance in the ejecta instead is connected mainly to the rapidity of loss of the envelope, so it is higher for higher initial mass, then increases again when the third dredge up begins to be effective.\\
For the composition of the models in display, the oxygen abundance in the ejecta is first correlated with the stellar mass, and decreases going down from 7 to 5.5\Msun. Decreasing the mass further, from 5.5 to 4.5\Msun,  another effect becomes dominant: the AGB phase lasts  longer (thanks to the lower luminosity and mass-loss rates), and the third dredge up episodes sensibly alter the envelope composition, by increasing the total CNO, because also $^{22}$Ne is dredged up from the helium intershell, and is immediately converted into sodium by p--captures. Thus the sodium abundance increases. As the total CNO is approximately constant in 2G stars, and sodium is not as large as predicted by models having M$\lesssim 4.5$\Msun, the 7--5\Msun\ range is the best for producing the 2G stars. In summary, the resulting composition of the ejecta is a complex result of the choice of convection model (temperature of HBB) and mass loss law.

\begin{figure*}
\vskip -35pt  
\begin{minipage}{0.98\textwidth}
\resizebox{1.\hsize}{!}{\includegraphics{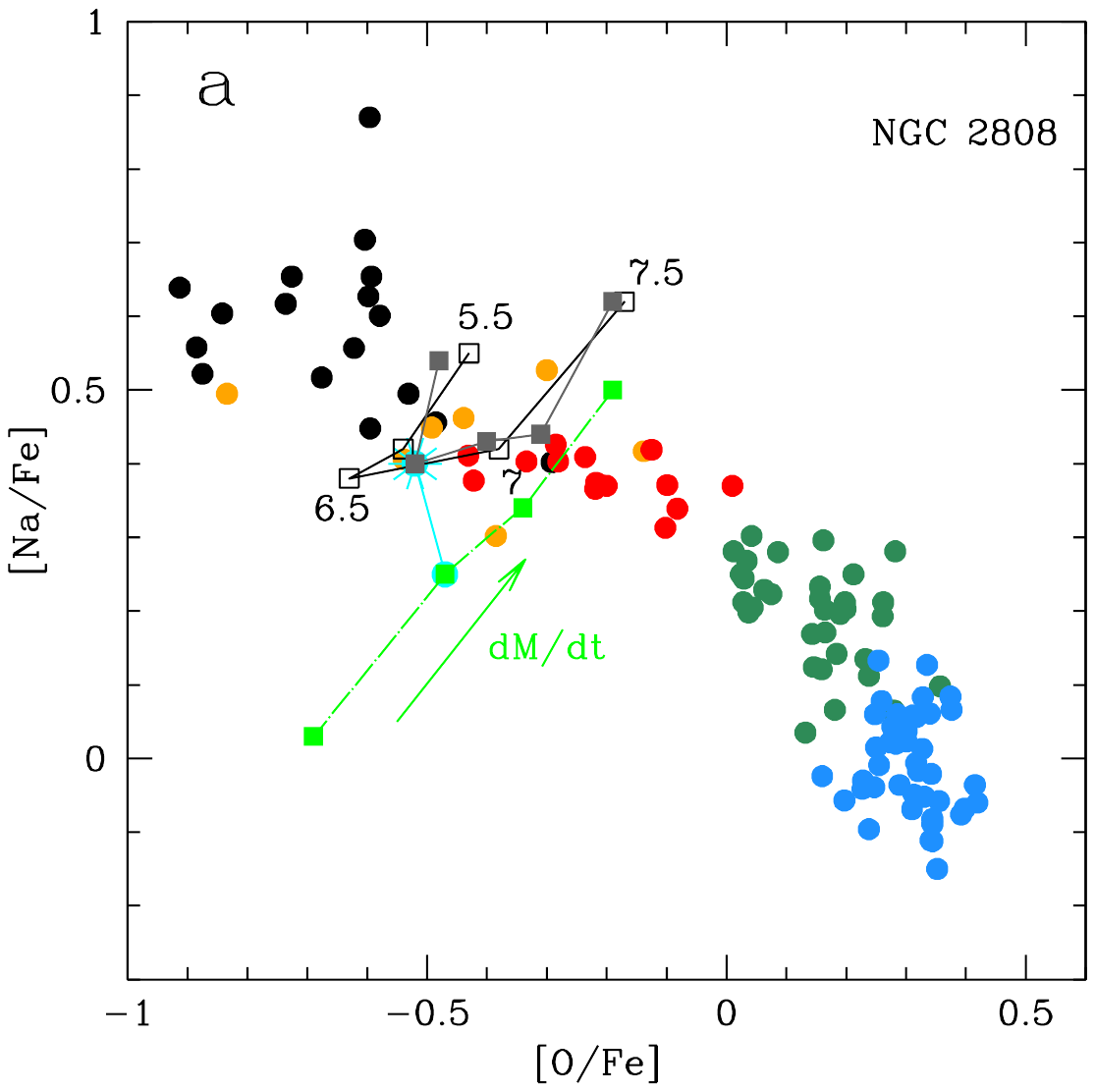}\includegraphics{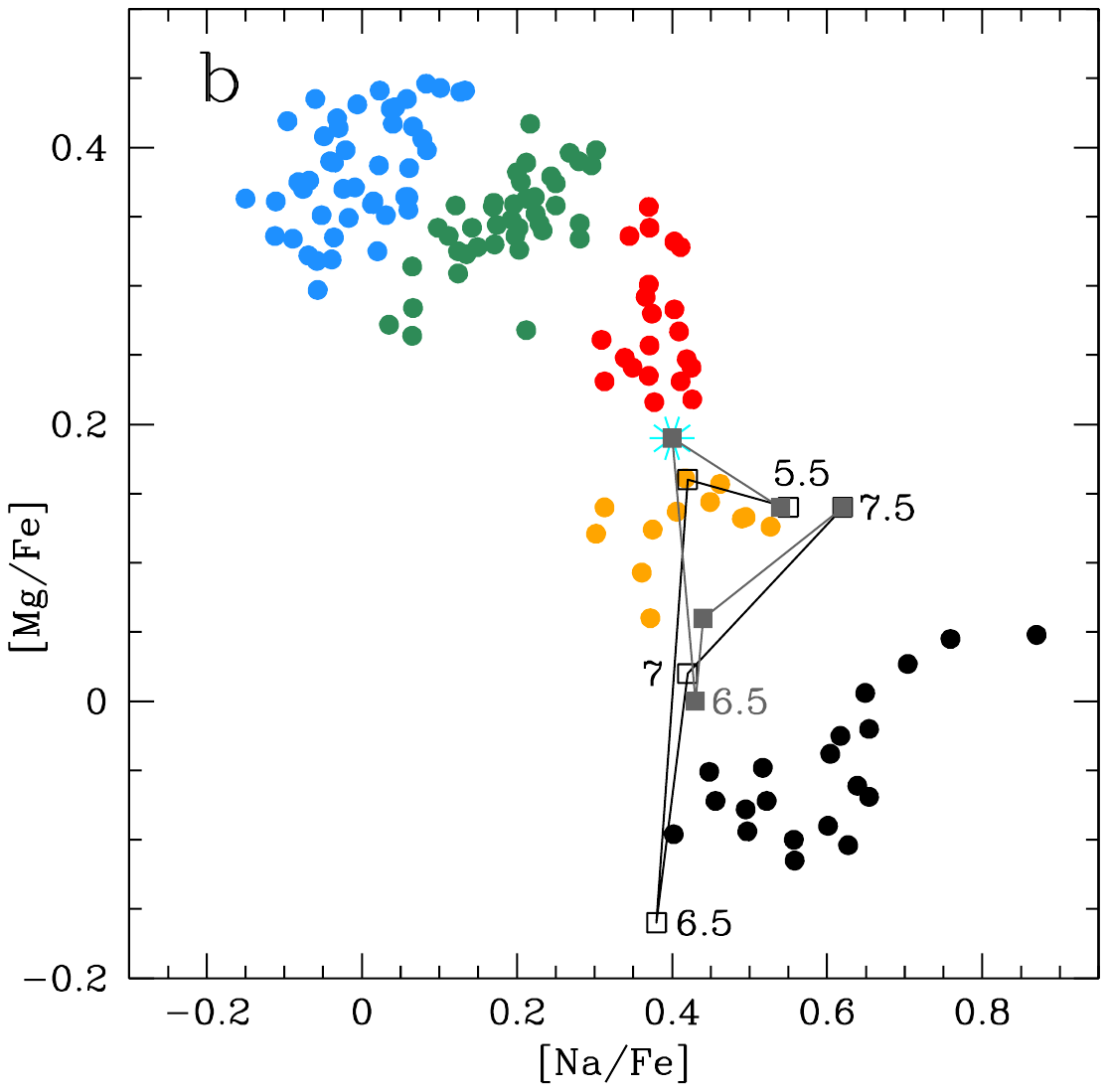}\includegraphics{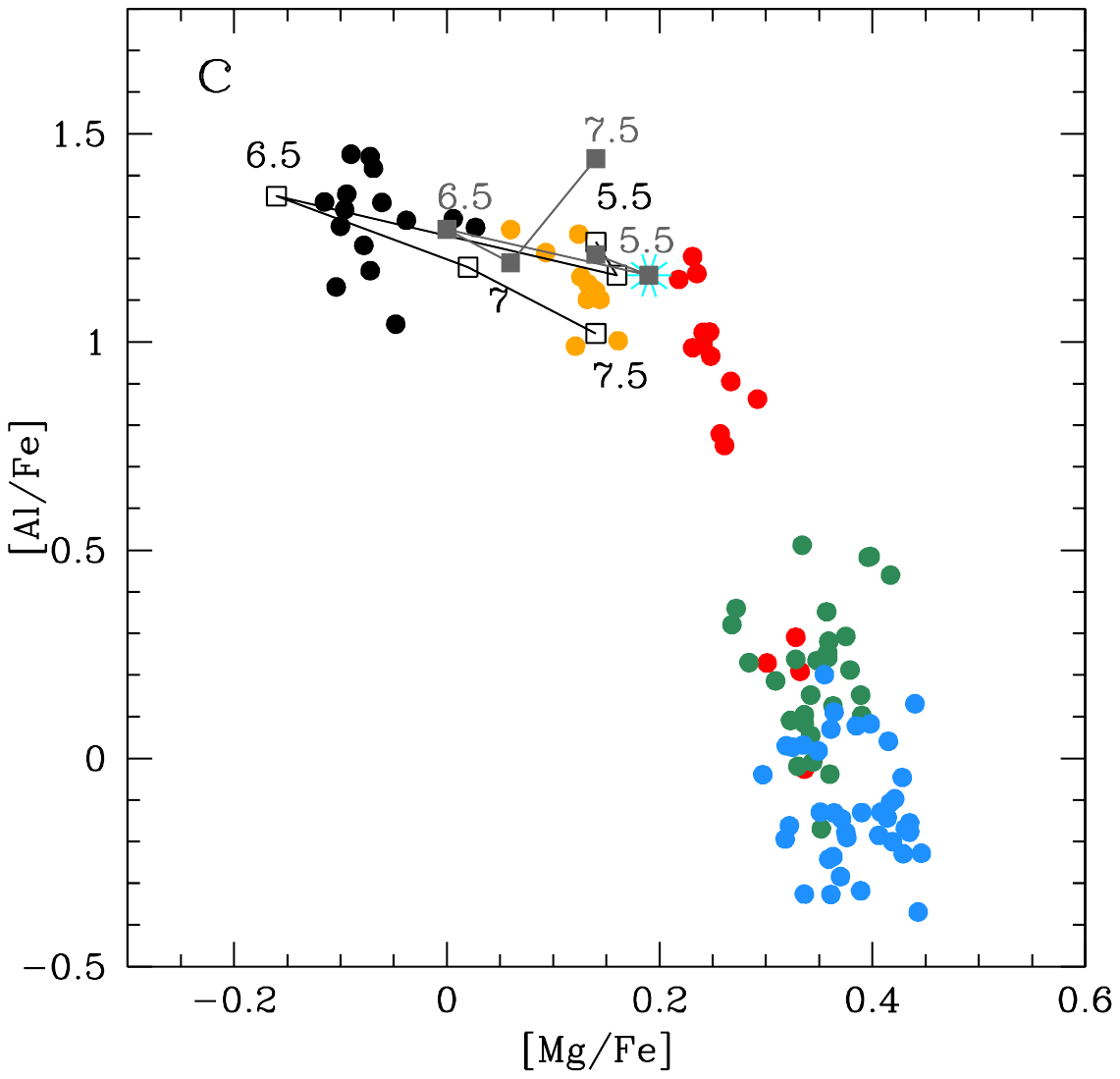}}	
\vskip -70pt
\resizebox{1.\hsize}{!}{\includegraphics{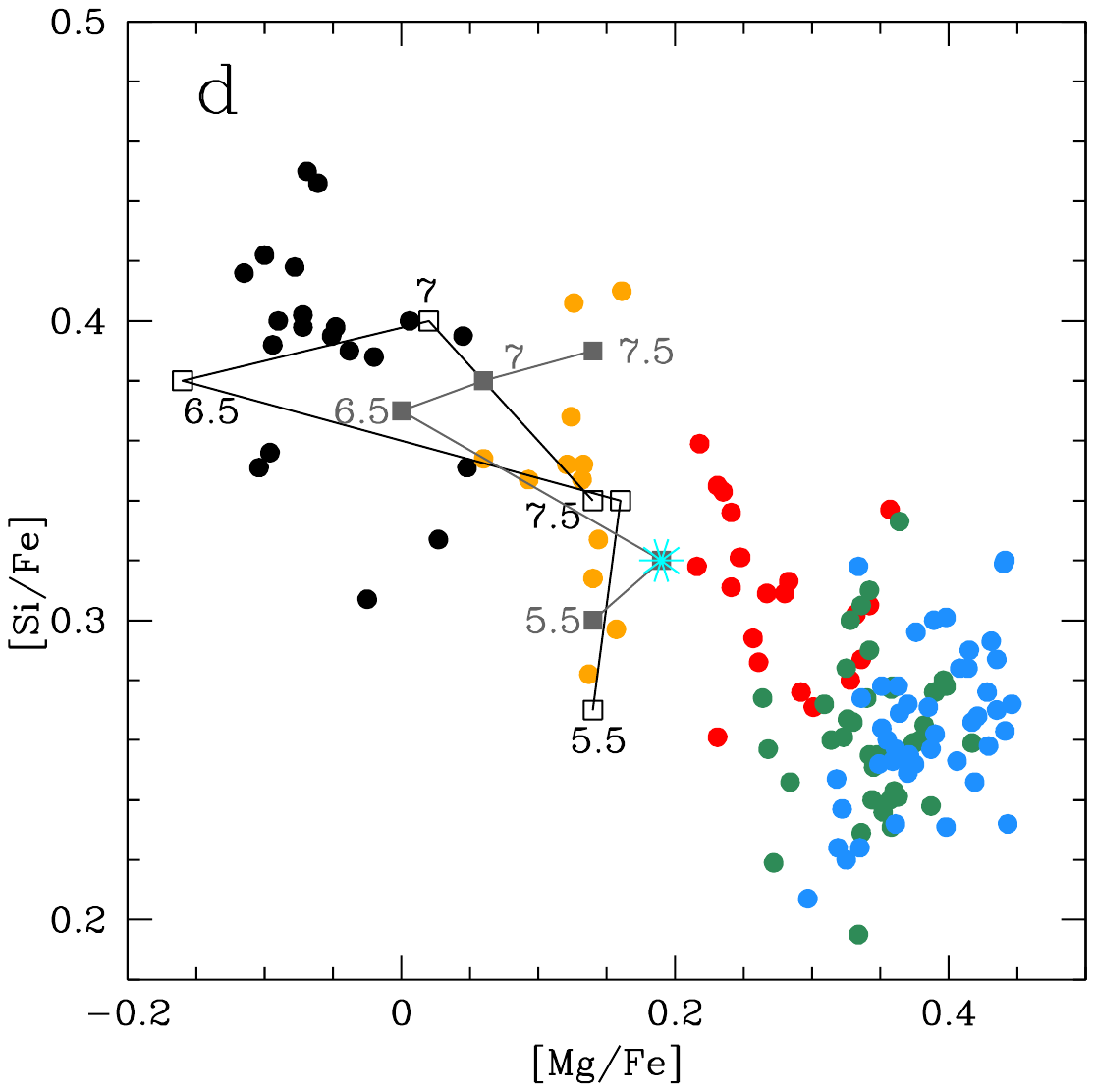}\includegraphics{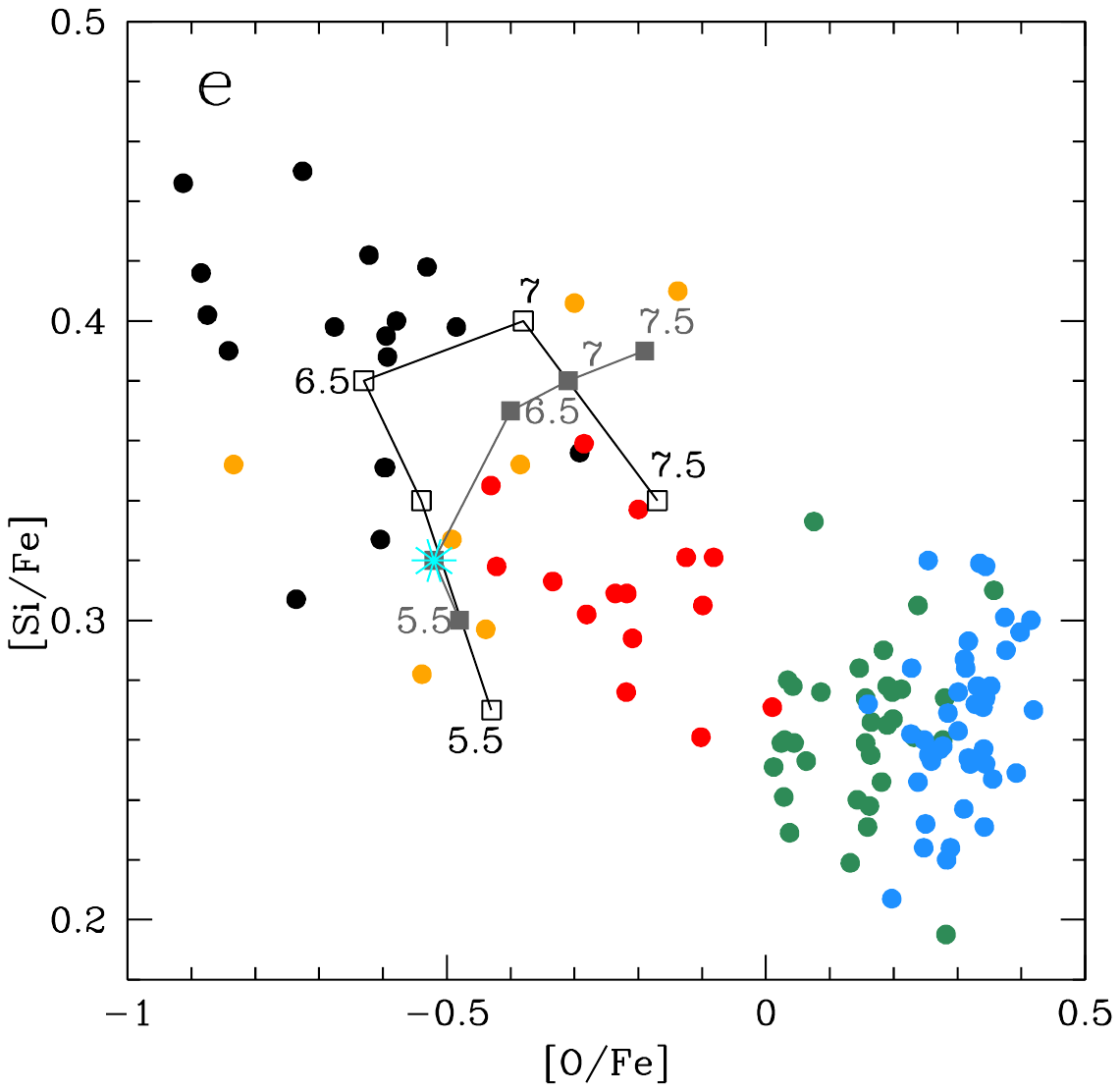}\includegraphics{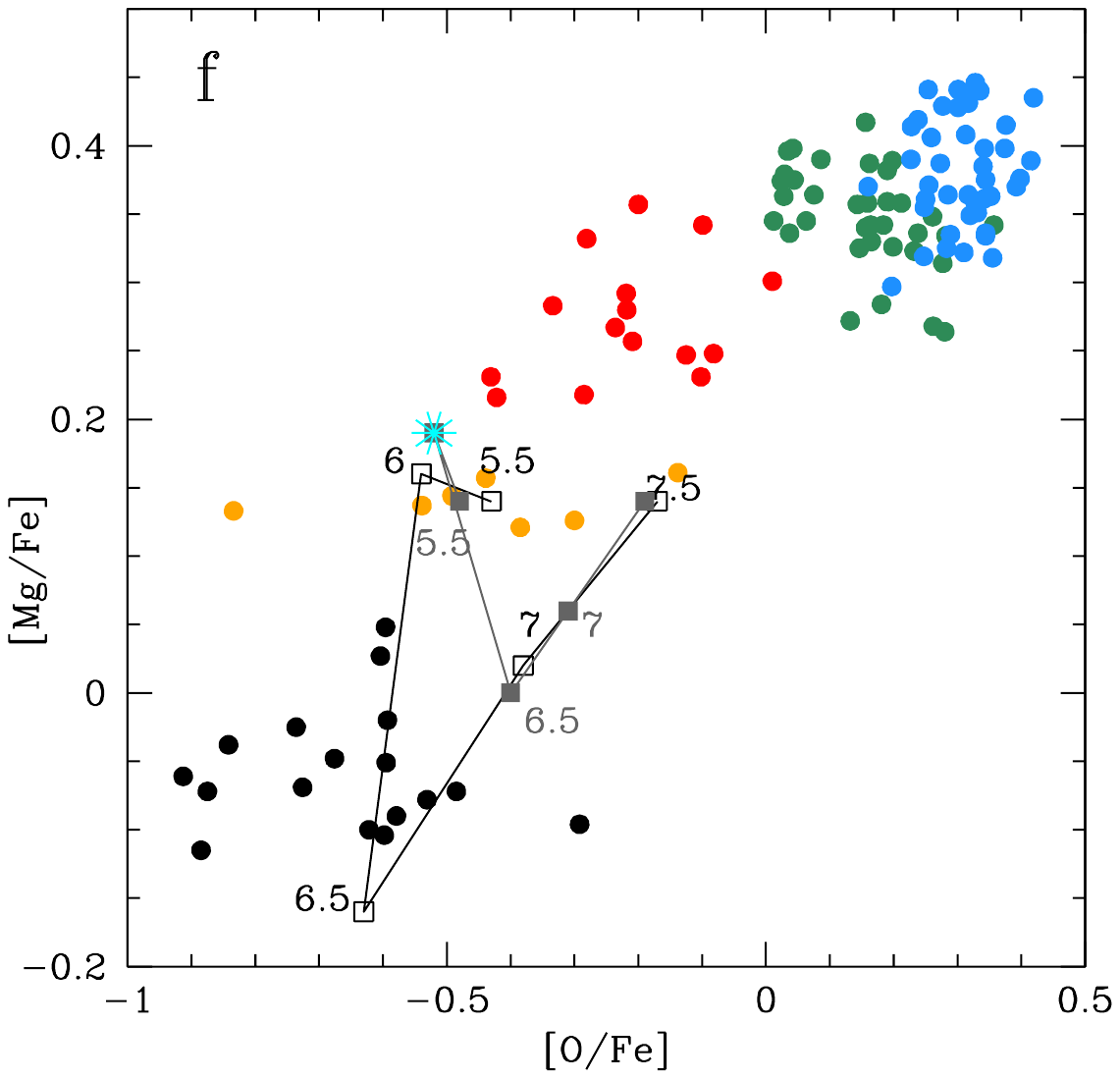}}	
\end{minipage}
\vskip-40pt
\caption{ The same as Fig.\,\ref{fig:3}, showing different model outputs. GREY: the squares represents the 5.5,  6.5, 7 and 7.5\Msun\ location for the new standard $\eta$\,0.25Ne2;  BLACK: (open black squares) models $\eta$\,0.25Ne2 for 5.5, 6, 6.5 and 7\Msun, for models having $\delta$[Fe/H]=--0.1.
The cyan big stars again defines the location of the new standard, $\eta$\,0.25Ne2, 6\Msun. In the panel a, the green squares are the 6\Msun\ ejecta for mass-loss rates with Reimers' parameter $\eta$\ going from $\eta$=0.001 (0.1 standard) to  $\eta$=0.02 (old standard), also shown in Fig.\,\ref{fig:3}.}
\label{fig:4}       
\end{figure*}

\subsection{How do we explain the extreme stars with oxygen abundances lower than allowed by models?}
\label{extramix}
The comparison (Figure\,\ref{fig:3}, panel a)  between the O--Na of the ejecta of the models (open blue circles) and the data for NGC\,2808 shows that the both the ``intermediate" I2 (orange dots) and the ``extreme" E stars (black dots) of the cluster have oxygen abundance up to 0.5--0.7\,dex smaller than modelled. As all these stars are giants, an easy escape from this discrepancy is met by remembering that the extreme abundances go together with ``pure" (undiluted) AGB ejecta, which also have a very high helium abundance \citep[Y$\sim$0.35--0.36,][]{ventura2010} are also more easily subject to extra mixing, as they develop a much smaller molecular weight barrier  \citep{dantona2007}, so these stars may lower their atmospheric [O/Fe] during the red giant branch evolution. The extra mixing depletes oxygen, but not sodium, so the O and Na will both fit the observed values. This escape anyway requires ad hoc modelling, if it is not limited to suggest an additional depletion by 0.1--0.2\,{\bf dex}, but instead implies a factor 2--3 reduction in the oxygen abundance. Further, and possibly more directly relevant, it hides a strong implication: we have to assume that all the I2 and E stars have been born from pure AGB ejecta, with helium content in the range Y=0.35-0.36. This may be consistent with the data for the cluster NGC\,2808: its color magnitude diagram shows the presence of a blue main sequence corresponding to a high helium content  \citep{dantona2005, piotto2007, milone2015}, and its ``extreme" population E shows spectroscopically the largest oxygen depletions, magnesium depletion and silicon enhancement of matter processed in very massive AGB or super--AGB stars  \citep{carretta2015, milone2017chromo, dantona2016}. On the other hand, for example, it was clearly shown  that  the maximum value allowed in the `extreme' stars of the cluster NGC 6402 is Y$\sim$0.31, although they show oxygen abundances corresponding to the composition of pure AGB ejecta \citep{dantona2022}. Also the study of the horizontal branch morphology \citep{tailo2020} or of the chromosome map \citep{milone2017chromo, milone2018he} limit the maximum helium of the 2G to Y=0.30--0.31 in several clusters, also for the stars which  have O--Na abundances expected for the pure ejecta of the models, and these have  much larger helium Y$\sim$0.35 \citep[e.g.][]{dantona2022}. The problem of discrepant helium abundances was first pointed out by \cite{bastian2015}  by considering the helium contents predicted by models for samples of stars with limited oxygen depletions. 
The way out of this conundrum requires that models  destroy more oxygen during the AGB evolution, so that the O--Na anticorrelation requires dilution with a larger quantity of pristine gas, lowering also the helium abundance in 2G stars \citep[see Fig.\,13 in][]{dantona2022}. 
Incidentally we note that a larger dilution may be in better agreement with the {\bf lithium} abundances in 2G stars, independently of the {\bf lithium} produced during the AGB evolution \citep{dantona2019}. 
Thus it is possible that extra--mixing in high helium stars is taking place, but its role must have a limited effect ($\sim$ 0.2\,dex) on the oxygen extra-depletion.\\
In regard to the other light elements, the problem is present only for the limited number of clusters which also show Mg depletion and Si increase: a better agreement for Mg and Si  requires longer timescales of evolution and HBB burning. So, in general, we need a longer AGB phase, that can be obtained by reducing the mass-loss rate, to fit better the O and Mg depletion and the Si increase. The undesired effect is to strongly deplete Na too, in disagreement with the observations.

In the following we discuss the status of art of the AGB models by our group published so far, concerning the other p-capture elements, and the other points of  weakness of these models.


\subsection{Magnesium depletion, Mg--Si anticorrelation and magnesium isotopes ratios}
In a fraction of clusters, also Mg shows abundance differences between the 1G and the 2G. In the AGB model, this implies that in some clusters HBB occurs at temperatures larger than in others.  Mg depletion is seen especially in low metallicity clusters, and this is consistent with the fact that HBB temperatures are larger in models with lower envelope opacities. An example of the trend of models and data with metallicity is provided in \cite{ventura2018} (see in particular their Fig.\,6). The best evidence is given by the observations of Mg abundances in \ocen\ \citep{meszaros2015, alvarezgaray2024}, where only the lower metallicity populations show Mg depletion. In this work  \cite{alvarezgaray2024} also show the presence in \ocen\ of a Mg--Si anticorrelation, indicating that  the Mg-Al cycle is extended to the $^{27}$Al(p,$\gamma$)$^{28}$Si reaction (see Fig.\,\ref{fig:1}). 
The ``standard" models built for NGC\,2808 do not account for the high spread shown by Mg and Si in this cluster (see the relevant panels in Fig.\,\ref{fig:3}).
\\

\subsection{The potassium versus magnesium anticorrelation and its extent}
\label{kmg}
The abundance of potassium in the 1G stars results from the nucleosynthesis in core collapse supernovae, whose uncertainties or metallicity dependence is still under investigation \citep[e.g.][]{kobayashi2011}. It was however surprising to find, among the red giants of the metal poor cluster NGC\,2419, a huge variation by a factor ten in its abundance, anticorrelated with magnesium abundance \citep{cohenkirby2012, mucciarelli2012}. 
As the magnesium depletion in the 2G stars implies that their gas had been processed at very high temperature ($\gtrsim$100\,MK), \cite{ventura2012ngc2419} proposed that  p--captures, at even larger temperatures, also manifacture potassium. In fact the cross sections of the chain  $^{36}$Ar(p,$\gamma$)$^{37}$K(e+,$\nu$)$^{37}$Cl(p,$\gamma$)$^{38}$Ar(p, $\gamma$)$^{39}$K 
increase by orders of magnitude at temperatures above 100\,MK. In {\bf super-AGB} models, \cite{ventura2012ngc2419}  were able to achieve  HBB temperatures \Thbb$\sim$120\,MK and K production was obtained by {\it increasing by a factor 100} the cross section of the reaction $^{38}$Ar(p,$\gamma$)$^{39}$K. \\
Presently, only three clusters\footnote{There is also a K spread in M\,54 \citep{carretta2022}, but it is not clearly anticorrelated with Mg or O.} are recognized to have a potassium spread anticorrelated with magnesium or oxygen: NGC\,2419, NGC\,2808 \citep{mucciarelli2015} and \ocen\ \citep{alvarezgaray2022}.\\
As we have pointed out for neon, also the abundances of the noble gas  $^{36}$Ar (an $\alpha$--nucleus that will be the most abundant form of argon at so low metallicities) and of the other argon isotopes  are not well known.

\section{An uncommon player: the initial neon abundance in the first generation gas}
\label{neonmassloss}
As we have reminded in Fig.\,\ref{fig:2nao}, the ``second dredge-up" occurring at the beginning of the AGB evolution affects not only the helium abundance \citep{ventura2010} but also the sodium and the neon isotopes processed in the interior. Fast p-captures on the dredged up neon contribute to increase sodium during the first phases of HBB \citep{ventura2005a, ventura2005b}. 

As hot bottom burning progresses, sodium is gradually destroyed. Therefore, the average sodium abundance in the stellar ejecta depends on the balance between the mass lost during the early phase (when sodium is enhanced due to second dredge-up and p-captures) and the mass lost later, when sodium has already been depleted.

In the meantime, the NO cycle continues to operate on the oxygen abundance which remains at the very low nuclear equilibrium values. For this reason, if we lengthen the AGB evolution slowing down the mass loss, we achieve a lower average oxygen, but also a lower average sodium abundance.\\
Based on the above discussion, it is not possible to reduce sufficiently oxygen without simultaneously destroying an excessive amount of sodium, assuming standard initial abundances. However, if the initial neon abundance is higher than assumed in standard compositions, the initially higher peak of dredged up neon converted into sodium helps to counteract the decrease in average sodium in the ejecta. Thus it is worth looking more in detail at models with increased initial neon.

\subsection{Role of the mass-loss rate and neon abundance}

In the top left panel of Fig.\,\ref{fig:3}, the ``standard" models ($\eta$=0.02) for [Fe/H]=--1.08  and for AGB masses from 7 to 4.5\Msun\ 
 are shown.  We take as reference the 6\Msun\ model, and show the abundances obtained by reducing $\eta$\ by a factor 2, 4 and 10: the abundances (green squares) shift to lower oxygen, and, obviously, to lower sodium, for the reasons outlined in Sect.\ref{whyneon}. Oxygen is depleted by a factor 10 in the model with the lowest mass-loss rate. Starting from this model location, we can recover a high sodium abundance if we {\it increase the initial neon abundance}: the blue square  and the big blue stars show the results obtained by multiplying the initial neon respectively by a factor two and four. The last model comes back to the range observed for  the range of sodium abundance. We identify specific models by their $\eta$\ reduction x and neon increase y as ``$\eta$\,x\,Ne\,y".
Increasing by a factor 2 and 4 the Ne abundance in the model with $\eta$=0.005, we reach the locations of the cyan big asterisk and cyan square.
Two ``dilution curves" (grey lines)  between the Na--O abundances in the models $\eta$\,0.1Ne4 or $\eta$\,0.25Ne2 and the initial abundances (big yellow star) are drawn. The triangles along the lines represent dilution of AGB ejecta with 20, 40, 60 and 80\% of pristine gas. 

\subsection{Discussion of the results: the most appealing compromise}

In figure \ref{fig:3} we compare the abundances of NGC\,2808 and the model predictions in the planes Na {\it vs.} O (panel a), Mg {\it vs.} Na (panel b), Al {\it vs.} Mg, (panel c) Si {\it vs.} Mg (panel d), Si {\it vs.} O (panel e), and Mg  {\it vs.} O (panel f),  with special attention both on the  $\eta$\,0.1Ne4 (big blue star) 6\Msun, and  on the $\eta$\,0.25Ne2 (cyan star) results. 
Panel a of Fig.\,\ref{fig:3} shows that the 6\Msun\ AGB ejecta of model  $\eta$\,0.1Ne4, diluted with pristine gas, can be representative of  the 2G stars in NGC\,2808, apart from very few E stars (black dots), for which we can still invoke some {\bf extra-mixing} during the giant branch evolution of these high--helium stars, as discussed in Sect.\,\ref{extramix}. 
Panel b shows that the Mg abundances of the I2 group are well reproduced by the same model, while the evolution of the 7\Msun\ is able to reproduce the Mg of the E group. A similar good agreement is found for Al (panel c) and Si (panel d). Unfortunately, the mass-loss rates in the models $\eta$\,0.1Ne4 are so low that already the 5\Msun\ evolution begin to be significantly affected  by the third dredge up. While, in fact, it fits very well (top red square) the Na--O abundances of the E stars,  its C+N+O abundance would be about 1.7 times larger than the initial CNO, at the limit of what is allowed from observations \citep{carretta2005}.\footnote{Only the clusters dubbed ``s--Fe anomalous" \citep{marino2015}  require star formation in gas polluted by AGB stars subject to important third dredge up \citep{dantona2016}.} In addition to the problem of CNO abundances, it is hard to invoke an increase of neon by a factor 4 (See Sect.\,\ref{PNNeon}).\\
The  $\eta$\,0.25Ne2 model (cyan star) of 6\Msun\  has Na and O compatible with the observed abundances. Oxygen is reduced by 0.8\,dex and leaves out only the extreme E stars.  
An increase by a factor two for neon is more reasonable, so  $\eta$\,0.25Ne2 is the best compromise for the 2G stars, {but it does not explain the extreme stars in group E.} \\

Panel b of Fig.~\ref{fig:3} shows that progressively reducing the mass-loss rate leads to lower abundances in the ejecta, as proton-capture reactions on both oxygen and magnesium have more time to operate. The increase in neon has a small or negligible effect on the final abundances. Anyway, the Mg depletion does not increase enough to reach the E sample location.  Lower mass-loss rates lead to represent better the Al production (panel c) although also for this element the abundances in the E stars is a bit larger, by $\sim$0.1\,dex, and is not reproduced even by the 7\Msun\ evolution. The discrepancy is anyway very small, considering that the range of variation in Al is$\sim$1.5\,dex. 
Similarly, both  Mg and O of group E are not explained (panel f), and the extension in Si abundances is not met (panels d and e) although the model location with lower mass-loss rates is largely improved.
\\
In conclusion, the position of the blue and cyan asterisks in the figures suggests that the location of the E group (blue asterisk) stars is met only by assuming a very low mass-loss rate (1/10 of the ``standard" rate adopted in all our previous models published so far) and a very high neon increase (4 times larger than the solar standard adopted abundance today). Both these choices are biased: 
a low mass-loss rate is compatible with the 2G chemical patterns only if the ejecta from stars near 5\Msun\ contribute very mildly to 2G formation. This significantly narrows the viable mass range to approximately $>$5--7\Msun\, thereby exacerbating the mass budget problem. Further, such a large neon abundance is not compatible with the observational data (see Sect.\,\ref{PNNeon}). The more modest increase in neon (2 times) and the more conservative reduction by a factor 5 in the mass-loss rate ( $\eta$=0.005) provides a good representation of the NGC\,2808 groups of abundances, apart from group E.\\
Therefore, an alternative mechanism must be operating and requires further investigation to fully account for the characteristics of class E.

\begin{figure}
\vskip -10pt  
\begin{minipage}{1.03\textwidth}
\vskip -15pt
\resizebox{.5\hsize}{!}
{\hskip -40pt \includegraphics{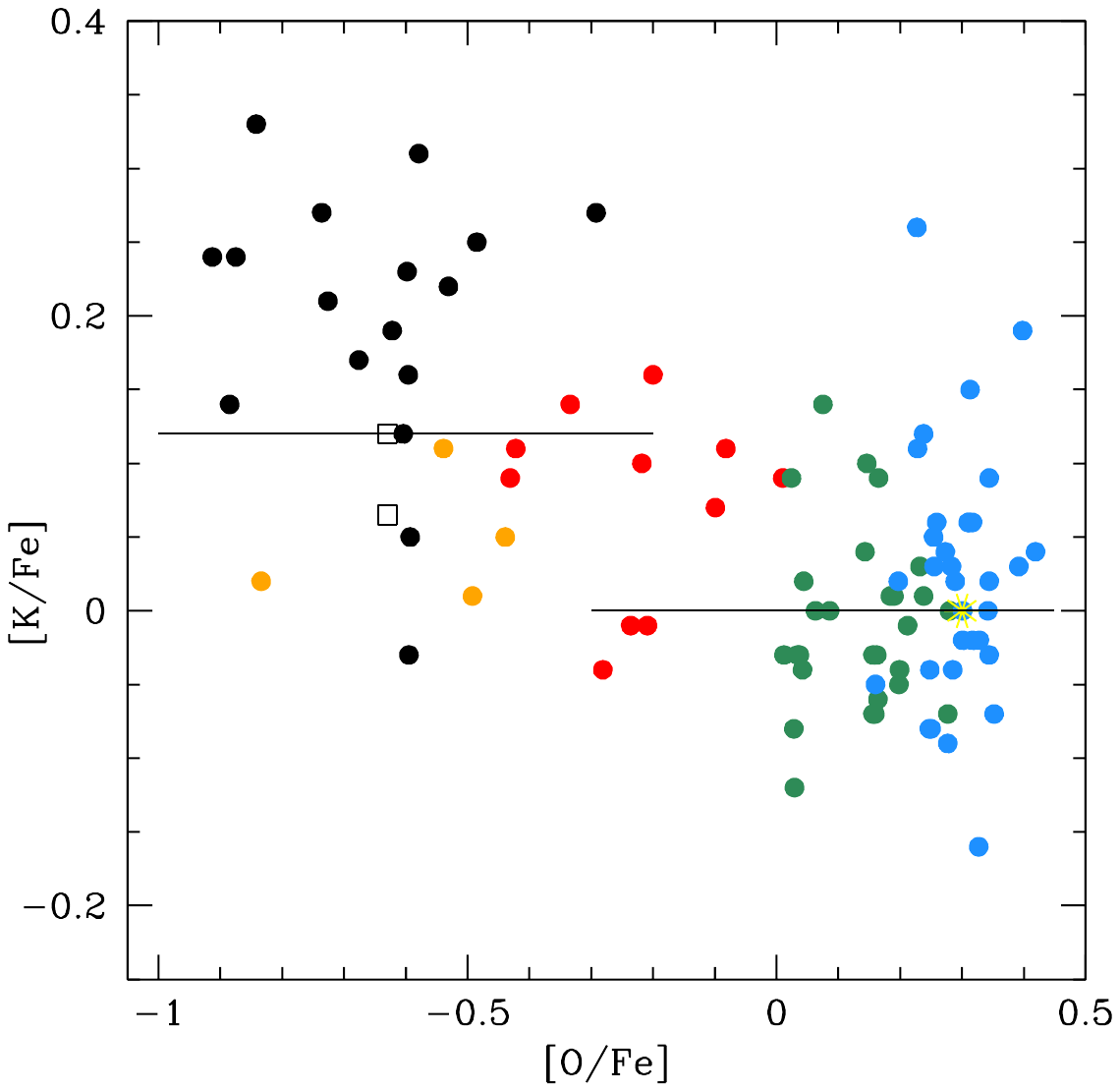}
\hskip -45pt
 \includegraphics{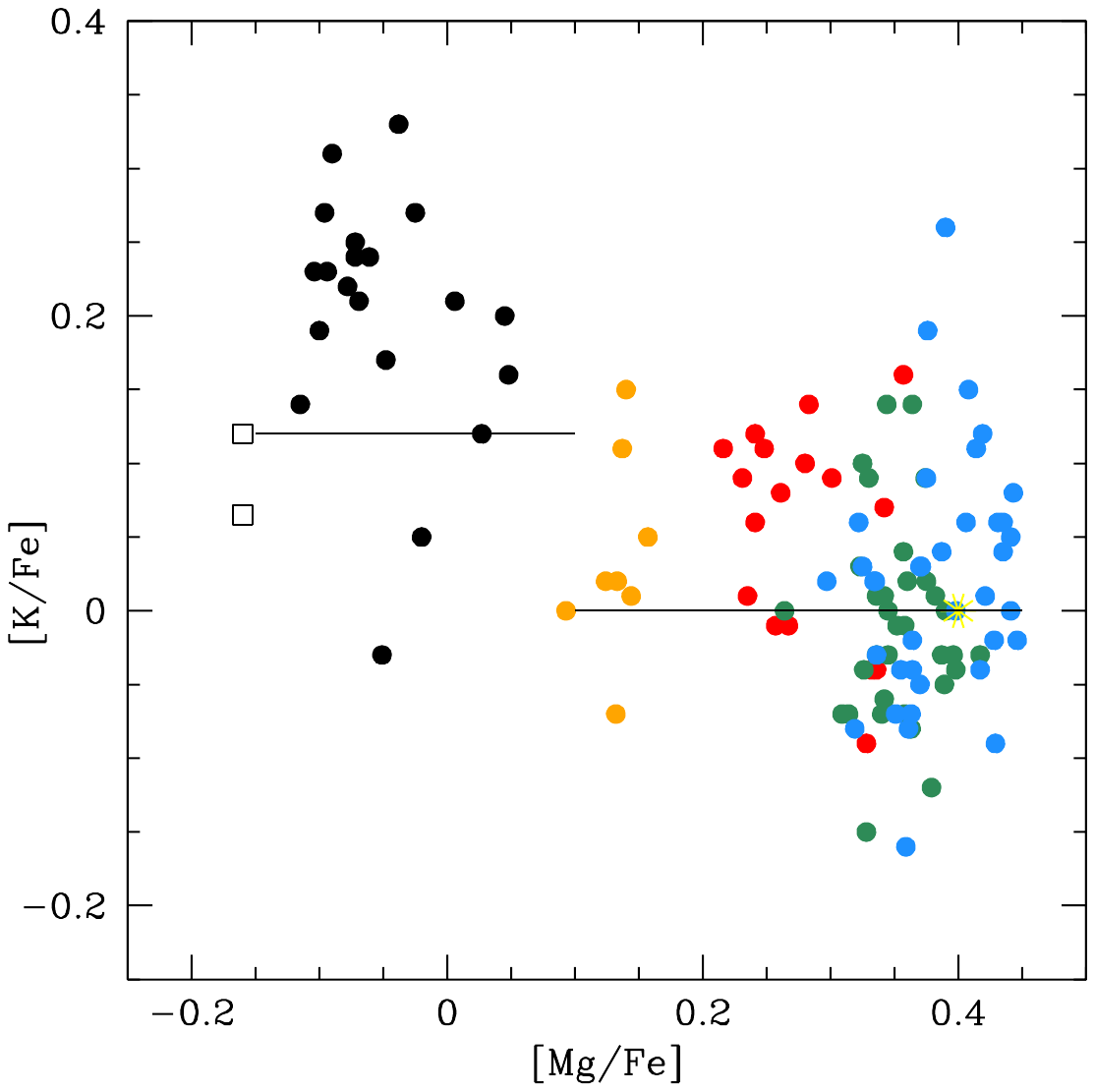} }
\end{minipage}
\vskip -40pt  
\caption{Potassium--oxygen and potassium--magnesium data for the GC NGC\,2808 (open circles) from \cite{carretta2015} and \cite{mucciarelli2015}. For all the symbols see Fig.\,\ref{fig:3}. Depletion for 6.5\Msun\ with [Fe/H] reduced by 0.1\,dex is shown by the open squares. The upper point ($\delta$[K/Fe]=0.125) is obtained by doubling the argon initial abundance. See text for further information.
}
\label{argon-pot}       
\end{figure}

\subsection{Models with reduced [Fe/H]}
\label{metals}
Fig.\,\ref{fig:4} shows as dark grey squares the average abundance in the ejecta of models for the case $\eta$\,0.25Ne2 and several different masses: 5.5, 6 (the standard case), 6.5, 7, 7.5\Msun. The abundances of the group I2 are better explained when we consider the role of different masses, while the E star are still problematic. \\
We therefore considered the abundances of ejecta for a chemical composition in which the iron content is reduced by $\delta$[Fe/H]=--0.1. Although in general the mass loss rate may be dependent on the metallicity, such a small change allows us to leave unchanged the mass loss parametrization at the value $\eta$=0.005. The results are plotted as open black squares and, interestingly, such a small abundance difference is able to shift the ``turning" mass (see Sect.\,\ref{oxygen}) from 6 to 6.5\Msun, so that larger elements depletions are achieved, and the abundances reach much closer to the location of group E stars in the different panels. The iron--reduced 6.5\Msun  performs better than the 6\Msun with standard iron content, and also provides results closer to the E group, for Al, Si and Mg, while it fails to fully reproduce O with a mild discrepancy (reduced to $\sim$0.2\,dex).

\section{A lower [Fe/H] for the E population? A proposal for the cluster formation}
\label{clumps}
We conclude that the E group abundances are consistent with the AGB model {\it if the AGB progenitors of this group have a slightly smaller metallicity than the bulk of other stars}. 
The F275W$-$F814W color spread of 1G stars in the pseudo two-color diagram—known as the chromosome map (ChM)—indicates that the 1G population of NGC 2808 is not chemically homogeneous. Instead, it exhibits a metallicity spread of approximately $\rm [Fe/H] \sim 0.1-0.2$ dex 
\citep{milone2015, dantona2016}. In particular, a comparison between the observed color distribution of 1G stars in NGC 2808 and synthetic colors from isochrones with varying metallicities suggests that the range between the 90th and 10th percentiles in the iron abundance distribution corresponds to $0.11 \pm 0.01$ dex \citep{legnardi2022}.

Supporting this, a differential analysis of high-resolution spectra for five stars spanning the 1G sequence on the ChM, conducted by \citet{lardo2022}, reveals a maximum iron variation of 
$0.25 \pm 0.06$ dex. These moderate iron variations had not been detected in earlier studies based on absolute spectroscopic measurements (e.g., Carretta 2015).
Thus such a iron spread was present at the formation of the 1G, or possibly it was introduced  while the cluster stars were already forming  \citep{lardo2023}. 

The formation of NGC\,2808 might have followed the root of hierarchical merging of different initial independent star forming clumps into a single massive cluster.  
This model is receiving attention in recent years, thanks to observations showing that young stellar systems have often a complex clumpy structure consisting of several stellar subsystems \citep[e.g.][]{kuhn2020, lim2020, dalessandro2021a, dalessandro2021b}. Some of these agglomerates may finally dissolve, but some of them may evolve into more massive stable clusters \citep[see, e.g.]{dalessandro2021b, dellacroce2023}, and, as shown by simulations, eventually hierachically merge into a single massive cluster \citep[e.g.]{livernois2021}. Attempts to link this merging to the multiple populations present in GCs --mainly concerning either age differences or the multiple iron abundance clusters-- are in \cite{gavagnin2016, hong2017, bailin2018}.  \\
We can assume that the population E is the second generation formed from the high mass AGB and super-AGB stars evolving in a massive clump which first formed stars. The other clumps started star formation later, and their gas was polluted by the core collapse supernovae exploding in the first clump \citep[e.g.][]{bailin2018}, so the stars formed in the other clumps were more metal rich. While the different clusters were merging into the final cluster NGC\,2808, the `first' 2G formed in the cooling flow of the high mass AGB of the first, more metal poor, cluster. To reproduce the abundances of the  group E, it is necessary that the difference in age between the age of the first clump and of the others be larger than about the time necessary for the 6.5\Msun\ to evolve, e.g. $\sim$60\,Myr, and it is also necessary that the first clumps are sufficiently massive that its pure AGB ejecta from $\sim$8 to $\sim$6.5\Msun\ are compatible with the mass of the population E.  When also the other clumps begin their AGB evolution, the winds from different evolving masses and different metallicities will merge in the cooling flow, and the metallicity will be, on average, larger.
The composition of the population E would then be that of pure ejecta, with high helium content Y=0.36--0.37, consistent with the location of the blue main sequence of NGC\,2808 \citep{dantona2005, piotto2007}.  
As the last supernova event occurs at $\sim$44\,Myr, for stars above 7.5\Msun,  we need an age difference of $\sim$15\,Myr between the birth of stars in the first clump and in the other ones.\\
After this period, the core-collapse supernova epoch will end in the other clumps, and the cooling flow onto the baricentre of the system will proceed including now a mixture of AGB ejecta from stars with different metallicities. The mixing timescale of the gas in the cooling flow will be short enough that the metallicity spread of these 2G stars will be very small, as observed \citep{legnardi2022}. Note that the mass of the clumps must be adequate to account for the present day population fraction. In particular, the stars in the population E are 14\%  \citep{carretta2015} of the total mass \citep[M=7.91$\times10^5$\Msun,][ and corresponding web page]{baumgardt2019}, so $\sim 1.1 \times 10^5$\Msun, must have all be born from the ejecta of the AGBs between $\sim$7.5 and $6.5-6$\Msun. For a standard Kroupa IMF the initial mass of the putative lower metallicity clump must then be $4-5\times 10^6$\Msun, so it is the most massive of the components.
We will examine the hypotheses in detail in a forthcoming dedicated paper. 
\\

\section{The Potassium problem}
\label{pot}
Fig.\,\ref{argon-pot} shows the last issue of chemistry in NGC\,2808, the abundances of potassium versus oxygen and potassium versus magnesium. More than an increasing trend of K for decreasing O or Mg, the data seem better represented by a single potassium abundance for all groups, apart from the E group, showing a K abundance higher by $\sim$0.2\,dex. 
We follow the idea of \cite{ventura2012ngc2419} and look at the effect of the chain $^{36}$Ar(p,$\gamma$)$^{37}$K(e+,$\nu$)$^{37}$Cl(p,$\gamma$)$^{38}$Ar(p, $\gamma$)$^{39}$K for the production of potassium.  Although 0.2\,dex is not a factor as huge as the 10 times increase in the cluster NGC\,2419 \citep{cohenkirby2012, mucciarelli2012}, the [Fe/H] of the present models is much larger, and the \Thbb's consequently smaller, so the AGB models do not show significant K increase. Nevertheless, for the highest masses of the models with [Fe/H] reduced by 0.1\,dex, and increasing the nuclear reaction rates as done for NGC\,2419, some small increase in K was obtained. 
The increase was by 0.065\,dex for the 6.5\Msun. The initial argon abundance was kept in the solar proportion. By doubling it, the total increase in K was by 0.12\,dex. We report this latter value in Fig.\,\ref{argon-pot}. The result is in the right direction but the requirements are so strong that it is not worth to try to achieve a better agreement. We only note that the argon abundance is a possible additional parameter to treat the problem of potassium.

\section{The neon abundances in Planetary Nebulae}
\label{PNNeon}
\begin{figure}
\begin{minipage}{0.48\textwidth}
\vskip -30pt  
\resizebox{1.\hsize}{!}{\includegraphics{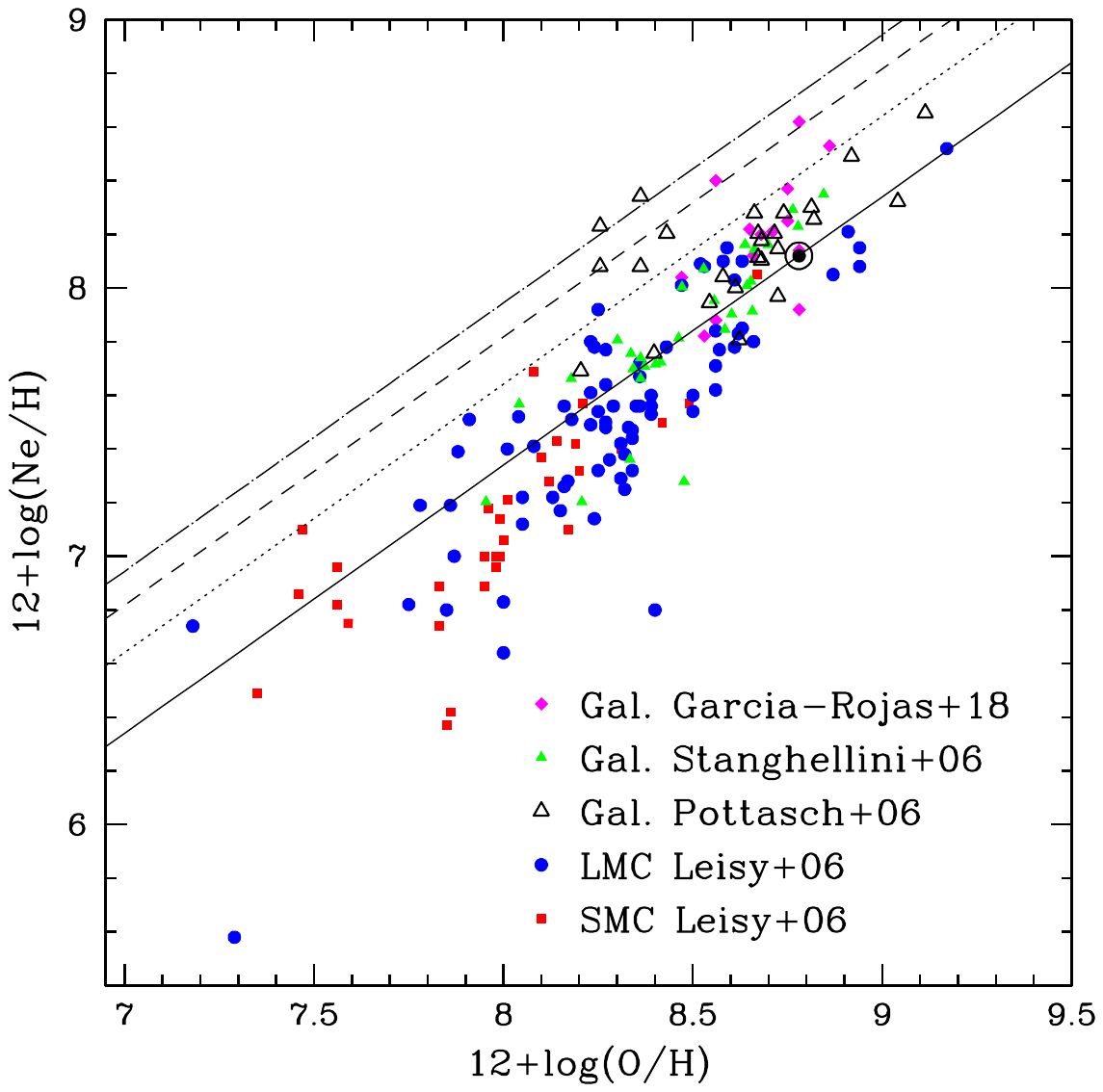} }	 
\end{minipage}
\vskip -70pt  
\caption{ Schematic neon versus oxygen abundances in planetary nebulae (PN). The solar point is located at the abundances from \cite{lodders2003}. The diagonal lines show the location for solar scaled abundances (full line) and for  neon abundances twice (dots), three times (dashed) and four times (dot-dashed) the solar value. Data are taken from different sources for PN, as labelled.}
\label{figne}       
\end{figure}

Fig.\ref{figne} shows the neon versus oxygen abundances in planetary nebulae (PN), where oxygen is a proxy for metallicity and the solar point is shown at the abundances by \cite{lodders2003}. The diagonal lines show the location for solar scaled abundances (full line) and in the hypothesis that the solar neon abundance is twice (dots), three times (dashed) and four times (dot-dashed) the solar value. Data are taken from different sources: galactic PN from \cite{garciarojas2018} (purple) and \cite{stanghellini2006} (green dots), galactic PN from \cite{pottasch2006}, Large  and Small  Magellanic Clouds PN \citep{leisy2006}. 
A detailed analysis of the data is beyond the scope of this work; however, it is worth noting that several measurements show neon abundances higher than solar at oxygen levels near solar, whereas only a few low-oxygen data points exhibit supersolar neon. Additionally, data at low metallicity are scarce. While this comparison remains inconclusive, it suggests that more neon abundance measurements are needed before drawing firm conclusions about the initial neon levels in low-metallicity environments, such as those found in proto-GCs.

\section{Conclusions}
In this work we make two proposals to soften the discrepancy between the AGB scenario and the distribution of the abundances of light elements among GC stars.

The first proposal is to consider whether the initial neon abundance of the forming cluster can be larger. 
As neon is an alpha element, an $\alpha$/Fe$\simeq$0.2--0.4 is generally already introduced in the choice of the initial neon abundance , we are now adding a further increase. We follow this approach because the solar abundance of noble gases is not well known, as summarised in Sect.\,\ref{intro}. Our comparisons are made with the data for GC NGC\,2808, one of the most complex for its extended light elements anti-correlations, very well studied in the spectroscopic sample by \cite{carretta2015}.
If we limit the initial neon increase to a factor two larger than the solar-scaled abundance, 
we may reduce by a factor four the mass-loss rate generally adopted in the models, and are able to explain better the anticorrelations, {\it apart from the abundance values of the most extreme group of the cluster}. 
The abundances in planetary nebulae (Sect.\,\ref{PNNeon}) do not contradict such a neon increase for solar metallicities, but there are too few data at the low metallicities of GCs to make a more significant comparison.
An alternative possibility is to consider the solar neon and argon abundances correct, and suggest that, at low metallicity, the ratios Ne/Fe and Ar/Fe are larger than what is implied by their being $\alpha$-elements. We then propose that an effort be made to improve our understanding of the stellar yields of noble gases and their galactic chemical evolution.
\\
Our second proposal is to suggest that the very high extension of the anticorrelations in the cluster NGC\,2808 is ultimately due to the coexistence in the cluster of different metallicities in the 1G stars. In the framework of the hierarchical cluster assembly scenario \citep[e.g.][]{dellacroce2023}, we show that the extreme stars in NGC\,2808 are better modeled {\it if the stars showing the most extreme anticorrelations have a metallicity smaller by 0.1\,dex}. As this cluster appears to have a global iron spread by $\sim$0.25\,dex  \citep{lardo2023} among its 1G stars, we suggest that it was formed by the assembly of multiple clumps, having slightly different metal abundances. \\
Finally, we try to model the potassium variations shown by the extreme stars, in the framework of the explanations proposed by \cite{ventura2012ngc2419} for the very metal poor massive GC NGC\,2419. The higher metallicity of NGC\,2808 works against the efficiency of the argon to potassium nuclear reaction chain, as the models \Thbb's are limited by the envelope opacities. Thus we are obliged to increase the cross sections of the relevant reactions by a factor 100, as done for the case of NGC\,2419, to achieve some production of potassium. We find, that increasing the initial argon abundances helps in achieving a larger potassium formation, but obviously this increase helps only if the nuclear reactions are active, so it does not soften the strong requirement of nuclear reaction rate increase.\\
In conclusion, we have shown  that the abundances of noble gases may have an important role in achieving a better match between the AGB model and the elemental abundances in the 2G stars in GCs, adding a different point of view on the necessity of knowing better the abundance of the neon (and argon) solar standard. 

\begin{acknowledgements}
We thank Eugenio Carretta, Ga\"el Buldgen and Emanuele Dalessandro  for useful discussions and informations on different relevant aspects of this work. We are also grateful to Eugenio Carretta for providing his data in tabular form.
\end{acknowledgements}

\bibliographystyle{aa}
\bibliography{noble}

\begin{appendix}
\section{Warnings on the use of results}
The results of the models computed for this project are given in Tables \ref{TAB1} and  \ref{TAB2}, as mass fractions (X$_i$) of the average abundance of the ejecta (Table\,\ref{TAB1}) or in the usual logarithmic form, with respect to the solar abundance ratios:
\begin{equation}
    [N_i/Fe] = {\log { 56 X_i\over A \times X_{Fe}} } -  {\log {56 X_{i,\odot}  \over A \times X_{Fe,\odot}}} 
\label{eq3}    
\end{equation}
in Table\,\ref{TAB2}.
We choose to provide extended tables, because the increase in the neon abundance, keeping the [Fe/H] at the same value, or the decrease of [Fe/H] by 0.1 and 0.2\,dex, alter the precise initial values of initial abundances, and may result in small differences also in the abundances of the ejecta that in principle should not be affected at all.\\ 
More importantly, we take the occasion to remember that the results given in the form of the Eq.\,\ref{eq3} in Table\,\ref{TAB2} can not be {\it arbitrarily scaled} in the logarithm planes generally adopted for the abundances, when they are compared with observations data sets. This misuse of the results happened a few times in the literature, even leading to a full misunderstanding, and thus we blame ourselves for not being explicit enough in clarifying slavishly this before.
A logarithm shift is allowed (with care) if the element in question is destroyed by p--capture reactions, because its percentual decrease will depend on the initial abundance. For instance, the oxygen and magnesium depletions depend on the initial abundances. In this case anyway it is important to remember and take into account the $\alpha$-enhancement adopted in the models. On the contrary, if an element is produced by a p-capture reaction involving (obviously) another independent abundance, `scaling'  is a meaningless procedure. For example, in the case of aluminium, the Al production depends on the initial Mg abundance and is totally independent from the initial Al abundance. If, e.g., the 1G stars of the cluster under study have an average [Al/Fe]=+0.2 (or [Al/Fe]=--0.2),  this does not mean that the values [Al/Fe] of col.\,11 of Table\,\ref{TAB2}  can be scaled up (or down) by 0.2\,dex, as they start from [Al/Fe]=0 and the increase is independent from the initial value.  Viceversa, if we look at [Mg/Fe] in col.\,10, and forget that the initial magnesium is 0.4\,dex larger than in the solar ratio, the positive values listed look like magnesium is {\it produced} in the models. 

\begin{table*}
\caption{Abundances of the relevant elements in the ejecta of the AGB models}
\centering 
\begin{tabular}{l c c c l c c c c c c c c}
M   & [$\alpha$/Fe] & f(Ne)  & Z(met) &  $\eta$(mloss) &      C  &       N   &      O &       Na &       Mg &  Al&        Si &       Fe  \\
\hline
& & & & initial & 2.11e-4 &  6.15e-5 &  1.28e-3 &  2.60e-6 &  1.92e-4 &  4.98e-6  & 1.49e-4 &  1.14e-4 \\  
\hline
6   & 0.4  &   1  &   0.0023  &   0.002  &     4.02e-5  & 1.34e-3 &  1.47e-4 &  2.78e-6 &  1.15e-4 &  5.85e-5 &  1.71e-4 &  1.14e-4 \\  
6   & 0.4  &   1  &   0.0023  &   0.005  &     4.61e-5  & 1.47e-3 &  2.10e-4 &  4.62e-6 &  1.23e-4 &  5.72e-5 &  1.67e-4 &  1.14e-4 \\    
6   & 0.4  &   1  &   0.0023  &   0.01    &     3.34e-5  & 1.11e-3 &  2.93e-4 &  5.69e-6 &  1.35e-4 &  4.87e-5 &  1.56e-4 &  1.14e-4 \\    
\hline
4.5 & 0.4  &   1 &  0.0023  &   0.02  &    9.89e-5 &  1.74e-3 &  5.01e-4 &  1.80e-5 &  1.35e-4 &  3.13e-5 &  1.49e-4 &  1.14e-4  \\
5   &  0.4 &    1 &  0.0023 &    0.02  &    1.64e-4 &  1.53e-3 &  4.37e-4 &  1.23e-5 &  1.27e-4 &  4.15e-5 &  1.50e-4 &  1.14e-4 \\ 
5.5 & 0.4 &    1 &  0.0023 &    0.02  &    4.62e-5 &  1.26e-3 &  3.75e-4 &  9.24e-6 &  1.27e-4 &  4.21e-5 &  1.51e-4 &  1.14e-4 \\
6    & 0.4 &    1 &  0.0023 &    0.02  &     3.05e-5 &  1.02e-3 &  4.16e-4 &  8.19e-6 &  1.29e-4 &  3.62e-5 &  1.52e-4 &  1.14e-4 \\
7    & 0.4 &    1 &  0.0023 &    0.02  &     3.08e-5 &  9.47e-4 &  5.13e-4 &  1.02e-5 &  1.34e-4 &  2.91e-5 &  1.52e-4 &  1.14e-4\\
\hline
6    & 0.4 &    2  &   0.0026 &   0.002 &      3.84e-5 &  1.51e-3 &  1.31e-4 &  5.07e-6 &  1.05e-4 &  7.54e-5 &  1.87e-4 &  1.14e-4 \\    
5.5  & 0.4 &   2 &    0.0026 &  0.005  &     6.00e-5 &  1.82e-3  &  2.11e-4 &  8.98e-6 &  1.16e-4 &  8.04e-5  & 1.66e-4 &  1.14e-4 \\    
6     & 0.4 &   2 &     0.0026 &  0.005 &      4.11e-5 &  1.31e-3  & 1.94e-4 &  6.54e-6 &  1.18e-4 &  7.20e-5  & 1.75e-4  & 1.14e-4 \\
6.5  & 0.4 &   2 &    0.0026 &    0.005 &      3.68e-5 & 1.18e-3  & 2.57e-4 &  7.06e-6  & 7.80e-5  & 9.26e-5 &  1.98e-4  & 1-14e-4  \\    
7    & 0.4  &   2  &   0.0026 &    0.005  &     3.97e-5 &   1.12e-3 &  3.11e-4 &  7.14e-6 &  8.90e-5 &  7.74e-5 &  2.03e-4 &  1.14e-4  \\   
7.5 & 0.4  &   2  &   0.0026 &    0.005   &    4.38e-5  & 1.07e-3 &  4.15e-4 &  1.09e-5  & 1.06e-4 &  5.51e-5 &  2.04e-4  & 1.14e-4   \\  
\hline 
6   & 0.4  &   4  &   0.0032  &   0.002  &   5.30e-5  & 1.62e-3 &  1.40e-4 &  7.50e-6  & 1.08e-4  & 7.54e-5 &  1.83e-4  & 1.14e-4 \\
6   & 0.4 &    4   &  0.0032 &    0.005   &    7.32e-5 &  1.69e-3 &  2.27e-4 &  9.66e-6 &  1.18e-4 &  7.36e-5&   1.75e-4 &  1.14e-4 \\    
5   & 0.4  &   4   &  0.0032  &   0.002  &   7.66e-5  & 2.39e-3 &  1.73e-4 &  1.22e-5  & 1.08e-4&   9.71e-5  & 1.71e-4 &  1.14e-4   \\ 
7   & 0.4  &   4  &   0.0032 &    0.002   &    4.62e-5 &  1.20e-3 &  1.93e-4 &  5.96e-6 &  6.21e-5 &  7.03e-5&   2.41e-4 &  1.14e-4 \\   
\hline
& & & & initial &  2.26e-4 &  6.62e-5 &  8.69e-4 &  2.79e-6 &  1.28e-4 &  5.32e-6 &   1.00e-4 &  1.14e-4     \\
6  &  0.2 &    4   &  0.0023 &    0.002  &     4.02e-5 &  1.15e-3 &  1.81e-4 &  6.39e-6 &  6.71e-5 &  5.20e-5  & 1.32e-4 &  1-14e-4 \\    
\hline
& & &$\delta$[Fe/H]=--0.2 & initial & 1.30e-4 &  3.82e-5&   7.94e-4&   1.62e-6&   1.18e-4&   3.07e-6&   9.18e-5&   6.95e-5 \\
6  &  0.4 &    2 &    0.0016&     0.005&     4.77e-5 &  1.42e-3 &   2.75e-5 &  3.71e-6 &   4.59e-5 &  5.99e-5 &  1.17e-4 &  6.95e-5 \\        
7  &  0.4 &    2 &    0.0016&     0.005 &     2.52e-5 &  8.10e-4 &  4.84e-5 &  2.46e-6&   1.70e-5 &  5.56e-5 &  1.58e-4 &  6.95e-5 \\   
\hline 
& & &$\delta$[Fe/H]=--0.1 & initial & 1.64e-4 &  4.83e-5 &  1.00e-3  & 2.04e-6  & 1.49e-4 &  3.88e-6 &  1.16e-4 &  8.77e-5\\
4.5 & 0.4  &   2  &   0.00202 &   0.005   &   1.15e-4 &  2.95e-3 &  3.76e-4 &  1.89e-5 &  1.20e-4  & 6.89e-5 &  1.23e-4 &  8.77e-5 \\   
5.0 & 0.4  &   2  &   0.00202 &   0.005  &     6.14e-5&   1.85e-3 &  1.80e-4 &  8.65e-6 &  9.27e-5 &  6.74e-5&   1.25e-4&   8.77e-5 \\                           
5.5 & 0.4  &   2  &   0.00202 &   0.005  &     6.50e-5 &  1.73e-3 &  1.86e-4&   7.26e-6 &  8.20e-5 &  6.74e-5  & 1.37e-4 &  8.77e-5 \\    
6  &  0.4  &   2   &  0.00202 &   0.005  &     4.13e-5 &  1.26e-3 &  1.45e-4 &  5.25e-6 &  8.60e-5 &  5.57e-5 &  1.42e-4 &  8.77e-5 \\    
6.5 & 0.4  &   2 &    0.00202 &   0.005  &     3.03e-5&   9.83e-4  & 1.16e-4&   4.93e-6 &  4.11e-5 &  8.80e-5 &  1.58e-4  & 8.77e-5 \\    
7  &  0.4  &   2  &   0.00202  &  0.005  &     3.25e-5  & 9.09e-4 &  2.09e-4 &  5.35e-6  & 6.25e-5  & 5.81e-5  & 1.65e-4 &   8.77e-5 \\    
7.5 & 0.4 &    2 &    0.00202 &   0.005 &      3.81e-5&   6.65e-4 &  3.38e-4 &  8.64e-6&   8.3e-5 &   4.09e-5 &  1.41e-4&   8.77e-5 \\    
7.5 & 0.4  &   2  &   0.00202 &   0.002 &      4.79e-5 &  9.09e-4&   2.43e-4  & 5.52e-6 &  6.87e-5 &  3.87e-5 &  1.90e-4  & 8.77e-5   \\   
\hline
\end{tabular}
\tablefoot{The quantities reported in the different columns are as follows. col.1: the mass in solar masses, col.2: the [$\alpha$/Fe] adopted for the distribution of abundances, col.3: the assumed neon abundance: f(Ne)=1 is the neon fraction corresponding to the $\alpha$-enhancement, while f(Ne)=2 and 4 correspond to a factor 2 and 4 more with respect to the $\alpha$-enhanced fraction;col.4: total metallicity; col.5: adopted value of the mass-loss parameter $\eta$\ in Eq.\,\ref{eq1}. The standard value adopted in the models  is $\eta$=0.02, so the standard models are at lines 4--8.  }
\label{TAB1}
\end{table*}

\begin{table*}
\caption{The same as in Table\,\ref{TAB1}, but in $\rm [X/Fe]$ units}
\centering 
\begin{tabular}{l c c c c c c c c c c c c}
M   & [$\alpha$/Fe] & f(Ne)  & Z(met) &  $\eta$(mloss) &    [C/Fe] & [N/Fe]& [O/Fe] & [Na/Fe] &[Mg/Fe] &[Al/Fe] &[Si/Fe] & age(Myr) \\     
\hline
& & & &  initial & 0.0 &   0.0 &   0.4 &   0.0 &    0.4 &    0.0 &    0.4  & \\
\hline
6   & 0.4  &  1  &  0.0023  &   0.002   & -0.72 &  1.34 & -0.54 &  0.03 &   0.18 &   1.07 &   0.46 &     70 \\
6   & 0.4  &  1  &  0.0023  &   0.005  &  -0.66  & 1.38 & -0.37 &  0.25 &   0.21 &   1.06 &   0.45 &     70 \\
6   & 0.4  &   1  &   0.0023  &   0.01    &   -0.80  & 1.26 & -0.01 &  0.34 &  0.25 &   0.99 &   0.42  &    70 \\
\hline
4.5 & 0.4  &   1 &  0.0023  &   0.02  &   -0.37 &  1.45 & -0.07 &  0.84 &   0.25 &   0.89 &   0.40 &    130 \\
5   &  0.4 &    1 &  0.0023 &    0.02 &    -0.12 &  1.40 & -0.07 &  0.67 &   0.22 &   0.92 &   0.40 &    100 \\
5.5 & 0.4 &    1 &  0.0023 &    0.02  &   -0.66 &  1.31 & -0.14 &  0.55 &   0.22 &   0.93 &   0.40 &    83 \\
6    & 0.4 &    1 &  0.0023 &    0.02  &   -0.84 &  1.22 & -0.09 &  0.50 &   0.22 &   0.86 &   0.41 &     70 \\
7    & 0.4 &    1 &  0.0023 &    0.02  &   -0.84 &  1.19 &  0.03 &  0.59 &   0.23 &   0.77 &   0.41 &     52 \\ 
\hline 
6    & 0.4 &    2  &   0.0026 &   0.002 &   -0.74  & 1.39 & -0.59 &  0.29 &   0.14 &   1.18 &   0.50 &     70 \\
5.5  & 0.4 &   2 &    0.0026 &  0.005  &  -0.54   & 1.47 &  -0.38 &  0.54&   0.14 &  1.21 &    0.45 &    83\\  
6     & 0.4 &   2 &     0.0026 &  0.005 &  -0.71  & 1.33 & -0.42  & 0.40  &  0.19  &  1.16 &   0.47 &     70 \\ 
6.5  & 0.4 &   2 &    0.0026 &    0.005 & -0.78  & 1.28 & -0.30  & 0.43  &  0.00  &  1.27 &   0.52 &     60 \\
7    & 0.4  &   2  &   0.0026 &    0.005  & -0.72 &  1.26 & -0.21 &  0.44  &  0.06  &  1.19 &   0.53 &    52  \\
7.5 & 0.4  &   2  &   0.0026 &    0.005   &-0.68 &   1.24 & -0.09 &  0.62 &   0.14 &   1.44&    0.54 &     45 \\
 \hline 
6   & 0.4  &   4  &   0.0032  &   0.002  &    -0.60&   1.42 & -0.56&   0.46 &   0.15&    1.18&    0.49 &     70 \\
6   & 0.4 &    4   &  0.0032 &    0.005   & -0.46  & 1.44&  -0.35&   0.57 &   0.19  &  1.17  &  0.47  &    70\\
5   & 0.4  &   4   &  0.0032  &   0.002  & -0.44  & 1.59 & -0.47 &  0.67  &  0.15  &  1.29  &  0.46   &  100 \\
7   & 0.4  &   4  &   0.0032 &    0.002   &  -0.66 &  1.29 & -0.42 &  0.36 &  -0.09 &   1.15 &   0.61 &     51\\
\hline
& & & &  initial & 0.0 &   0.0 &   0.2 &   0.0 &    0.2 &    0.0 &    0.2  & \\
6  &  0.2 &    4   &  0.0023 &    0.002  &   -0.75 &  1.24 & -0.88 &  0.36  & -0.08  &  0.99  &  0.32   &   70 \\
\hline
6  &  0.4 &    2 &    0.0016&     0.005&  -0.44 &  1.57&  -0.66 &  0.36 &  -0.01  &  1.29 &   0.50  & 70\\  
7  &  0.4 &    2 &    0.0016&     0.005 & -0.71  & 1.33 & -0.82  & 0.22 &  -0.42 &   1.26 &   0.64 & 51  \\
\\
\hline
& & &$\delta$[Fe/H]=--0.1 &&&&&&&\\
4.5 & 0.4  &   2  &   0.00202 &   0.005   &    -0.15 &  1.78 & -0.02 &  0.97  &  0.30&    1.25 &   0.42  &   129  \\
5.0 & 0.4  &   2  &   0.00202 &   0.005  &   -0.42 &  1.58 & -0.35 &  0.63  &  0.18 &   1.24 &   0.43 &    100 \\
5.5 & 0.4  &   2  &   0.00202 &   0.005  &  -0.40 &  1.55 & -0.33 &  0.55 &   0.14 &   1.24  &  0.42  &    83 \\
6  &  0.4  &   2   &  0.00202 &   0.005  &   -0.60 &  1.42 & -0.44 &  0.42 &   0.16  &  1.16  &  0.49 &     69  \\
6.5 & 0.4  &   2 &    0.00202 &   0.005  & -0.73  & 1.31&  -0.53 &  0.38 &  -0.16  &  1.35  &  0.53   &   59 \\
 7  &  0.4  &   2  &   0.00202  &  0.005  &  -0.71  & 1.27&  -0.28 &  0.42 &   0.02 &   1.18  &  0.55  &    51 \\
7.5 & 0.4 &    2 &    0.00202 &   0.005 &   -0.63  & 1.14 & -0.07&   0.62 &   0.14 &   1.02  &  0.49  &    45 \\
7.5 & 0.4  &   2  &   0.00202 &   0.002 &  -0.53  & 1.27 & -0.21  & 0.43 &   0.07 &   0.99 &   0.61   &   45 \\
\hline
\end{tabular}
\label{TAB2}
\end{table*}
\end{appendix}

\end{document}